\documentclass[12pt,preprint]{aastex}
 \usepackage{url}

\shorttitle{Magnetar as Central Engine of Gamma-Ray Bursts}
\shortauthors{Zou et al}
\slugcomment{}

\usepackage{lineno}
\begin{document}

\title{Magnetar as Central Engine of Gamma-Ray Bursts: Central Engine-Jet Connection,
Wind-Jet Energy Partition, and Origin of Some Ultra-Long Bursts}

\author{Le Zou, Zi-Min Zhou, Lang Xie, Lu-Lu Zhang, Hou-Jun L\"{u}, Shu-Qing Zhong, Zhen-Jie Wang, and En-Wei Liang$^{\ast}$}
\affil{Guangxi Key Laboratory for Relativistic Astrophysics, School of Physical Science and Technology, Guangxi University, Nanning 530004, China; lew@gxu.edu.cn}

\begin{abstract}
Gamma-ray burst (GRB) central engines and jet production mechanisms are still open questions. Assuming that the shallow decay segments of canonical X-ray afterglow lightcurves of {\em Swift} GRBs are attributed to the magnetic dipole (MD) radiations of newly-born magnetars, we derive the parameters of the magnetars and explore their possible relations to jet and MD wind emission. We show that the magnetar initial spin period ($P_0$) are tightly correlated with the jet energy ($E_{\rm jet}$), which is almost proportional to the wind energy ($E_{\rm wind}$). Our least square fits yield $P_0\propto E^{-0.36\pm 0.03}_{\rm jet}$ and $E_{\rm wind}\propto E^{0.91\pm 0.07}_{\rm jet}$. These relations may imply that a magnetar with faster rotating speed can power a more energetic GRB, and energy partition between the jet and wind may be quasi-universal. Although the $P_0-E_{\rm jet}$ relation is driven by a few sub-energetic GRBs in our sample, our Monte Carlo simulation analysis shows that sample selection biases from instrumental flux limits and contaminations of the bright jet afterglows cannot make this correlation. Within this jet-wind paradigm, we propose that GRB 101225A-like ultra-long GRBs, whose prompt gamma-ray/X-ray lightcurves are featured as a long-lasting plateau with a sharp drop, may be the orphan MD wind emission being due to misalignment of their jet axis to the light of sight. Brief discussion on the orphan MD wind emission and its association with the gravitational wave radiation of newly-born magnetars is presented.

\end{abstract}

\section{Introduction}
It is well believed that gamma-ray bursts (GRBs) and their afterglows are from a relativistic ejecta with an initial Lorentz factor $\Gamma_0\gtrsim 100$ based on the observed GRB spectra (Pacz{\'y}ski 1986; Goodman 1986; Rees \& M\'{e}sz\'{a}ros 1992; M\'{e}sz\'{a}ros \& Rees 1993; Woods \& Loeb 1995; Lithwick \& Sari 2001; see Kumar \& Zhang 2015 for a recent review). If the ejecta is beamed, its afterglow lightcurve should be a broken power-law whose decay slope\footnote{Notation $F_{\rm \nu}\propto \nu^{-\beta}t^{-\alpha}$ is adopted.} may change from $\sim 1$ to $\sim 2$ once the Lorentz factor of the ejecta being smaller than $1/\theta_j$, where $\theta_j$ is the jet opening angle (Rhoads 1999; Sari et al. 1999). Such a feature was detected in some optical afterglow lightcurves, favoring the GRB jet model (e.g., Harrison et al. 1999; see Frail et al. 2001; Liang et al. 2005, 2008; Wang et al. 2018 for sample analysis). Models of central engines for powering such a relativistic jet are sorted into two groups. One is a hyper-accreting stellar-mass black hole (e.g., Popham et al. 1999; Narayan et al. 2001; Lei et al. 2013), which may power a relativistic jet via $\nu \bar{\nu}$ annihilation in a neutrino-dominated accretion flow (NDAF; Ruffert et al. 1997; Popham et al. 1999; Chen \& Beloborodov 2007; Lei et al. 2009; Liu et al. 2017) or Blandford-Znajek mechanism for tapping the spin energy of a black hole (Blandford \& Znajek 1977; Lee et al. 2000; Li 2000). Another one is a rapidly spinning, strongly magnetized neutron star (a millisecond magnetar; Usov 1992; Duncan \& Thompson 1992; Thompson 1994; Dai \& Lu 1998; Wheeler et al. 2000; Ruderman et al. 2000; Zhang \& M\'{e}sz\'{a}ros 2001; Dai et al. 2006; Zhang \& Dai 2008, 2009; Metzger et al. 2008, 2011; Bucciantini et al. 2012; Metzger \& Piro 2014). A newly-born magnetar may provide enough rotational energy to power an ejecta via accreting the surrounding matters (Usov 1992; Duncan \& Thompson 1992; Metzger et al. 2011), and the residual rotational energy may be lost via magnetic dipole (MD) wind and gravitational wave (GW) radiations (Zhang \& M\'{e}sz\'{a}ros 2001; Lasky \& Glampedakis 2016; L\"{u} et al. 2018). Xu \& Huang (2015) proposed that the spin-down of the magnetar experiences two stages that correspond to the prompt gamma-ray phase and the afterglow phase, being due to the decrease of the tilt angle of the magnetic field.

The injected kinetic luminosity to the MD wind from the spin-down of a magnetar evolves as $L_{\rm k}\propto (1+t/\tau)^{-\alpha}$, where $\tau$ is the characteristic spin-down timescales of the magnetar (e.g., Zhang \& M\'{e}sz\'{a}ros 2001). The $\alpha$ value depends on the spin-down energy lost via the MD wind or the GW radiations, i.e., $\alpha=1$ in case of that the rational energy lost is dominated by the GW radiations, $\alpha=2$ if the MD radiation dominated, and $\alpha\in \{1, 2\}$ in a generic scenario for a stable magnetar. Transition from the GW to MD radiation dominated epoch may show up a smooth break with slope changing from $1$ to $2$ (e.g., Zhang \& M\'{e}sz\'{a}ros 2001; L\"{u} et al. 2018). The early $L_{\rm k}$ injection behaviors as a shallow decay phase at $t<\tau$. These features, especially the shallow decay phase, would be essential for probing the nature of GRB central engines.

The {\em Swift}/X-ray telescope (XRT) has made extensive observations to GRBs triggered by the Burst Alert Telescope (BAT) and collected a large sample of X-ray afterglow data with a temporal coverage from tens or hundreds of seconds to hours even days post the BAT trigger. A canonical XRT lightcurve that composes of several power-law segments and erratic flares was proposed (Zhang et al. 2006; Nousek et al. 2006). The most striking feature of the canonical XRT lightcurve is its shallow-to-normal decaying segment. It is apparently consistent with the $L_k$ injection behavior of the DM wind from a magnetar (e.g., Zhang et al. 2006; Liang et al. 2007). It would signal a magnetar as the central engines of these GRBs (Zhang et al. 2006; L\"{u} \& Zhang 2014; L\"{u} et al. 2015, 2018). Interestingly, an X-ray plateau followed by a very sharp drop is convincingly observed in GRB 070110 (Troja et al. 2007) and is plausibly detected in other GRBs (e.g., Liang et al. 2007; Lyons et al. 2010; Rowlinson et al. 2013; L\"{u} \& Zhang 2014; L\"{u} et al. 2015). The X-ray plateau would be internal energy dissipation of the MD wind, but have nothing to do with the external shocks. The rapid flux drop of the plateau indicates that the $L_k$ injection of the MD wind is suddenly shut down, likely suggesting that the magnetar collapses to a black hole before $\tau$ without further spin-down energy injection to the wind (Troja et al. 2007). Constraints on the physical properties of the magnetars in these GRBs with their XRT data have been presented by some authors (e.g., L\"{u} \& Zhang 2014; L\"{u} et al. 2015, 2018). Relations of these properties to the jet and MD wind radiations may give insights to the jet and MD wind productions.

Ultra-long GRBs with a duration longer than $10^4$ seconds are proposed as a unique population from different progenitors (Gendre et al. 2013; Stratta et al. 2013; Virgili et al. 2013; Levan et al. 2014; Ioka et al. 2016; Hou et al. 2018). However, the observed extreme difference between the jet and MD wind radiations indicates that viewing angle are substantial for identifying a burst in the jet and wind co-existing system. Both the prompt and afterglow radiations of the jet as well as the MD radiations could be simultaneously observed for on-axis observation to the jet. The jet emission might be missed and only the MD radiations could be observed for an off-axis observer since the MD wind should be quasi-isotropic (e.g., Metzger \& Berger et al. 2012; Gao et al. 2013). Off-axis observations to bright and soft gamma-rays/X-rays from the MD wind may be misidentified as a new population of ultra-long GRBs. The prompt gamma-ray/X-ray emission of ultra-long GRBs 101225A and 170714A are steady and long-lasting (Campana et al. 2011; Th\"{o}ne et al. 2011; Hou et al. 2018). It is interesting whether they are off-axis observations to a typical GRB and their prompt emission is dominated by the MD wind of magnetars.

In this paper, we assume that the shallow-decay segment observed in the XRT lightcurves is attributed to the MD wind emission and derive the parameters of the magnetars, and search for possible connections to the energies release of the jet and wind. In this jet-wind co-existing picture, we also explore whether some ultra-long GRBs are due to off-axis viewing effect to an energetic jet-wind system. Our sample selection and data analysis are presented in \S2. Properties of the magnetars and their connections with jet emission are shown in \S3. We explore the jet-wind connection and their energy partition in \S 4, and investigate whether ultra-long GRBs 101225A and 170714A like GRBs are due to the off-axis observations to a typical GRBs powered by a newly-born magnetar in \S 5. Discussion on observational biases and orphan MD wind emission is presented in \S 6. We summary our results in \S7. Throughout the paper, a concordance cosmology with parameters $H_0 = 70$ km s$^{-1}$ Mpc $^{-1}$, $\Omega_M=0.3$, and $\Omega_{\Lambda}=0.7$ is adopted.

\section{Sample Selection and Data Analysis}
The GRBs in our sample are selected from the current {\em Swift} GRB catalog. Their BAT and XRT data are downloaded from the web site of the {\em Swift} burst analyser (Evans et al. 2010)\footnote{http://www.swift.ac.uk/burst\_analyser/}. In order to make a joint X-ray lightcurve in the XRT band (0.3-10 keV) from the BAT trigger time to late epoch, the lightcurve of the prompt X-ray emission of a GRB is derived by extrapolated the BAT spectrum to the XRT band (O'Brien et al. 2006; Evans et al. 2007; Evans et al. 2009). Two apparently different types of XRT lightcurves are observed in the current Swift GRB sample. One is canonical XRT lightcurves (Zhang et al. 2006; Nouseck et al. 2006), and the others are monotonously decay as a single power-law (SPL) function (Liang et al. 2009). The canonical XRT lightcurves are characterized with an early shallow-decay segment or a plateau, which is attributed to the MD emission of a newly-born magnetar as mentioned in \S1. We select only the GRBs that have a canonical XRT lightcurve. Discussion on this sample selection effect on our analysis is presented in \S 6. Since the decay slope predicted by standard external shock models is steeper that 0.75 (e.g., Liang et al. 2007, 2008), we adopt a criterion of $\alpha<0.75$ for selecting a shallow-decay segment.

\begin{figure}
\includegraphics[angle=0,scale=0.5]{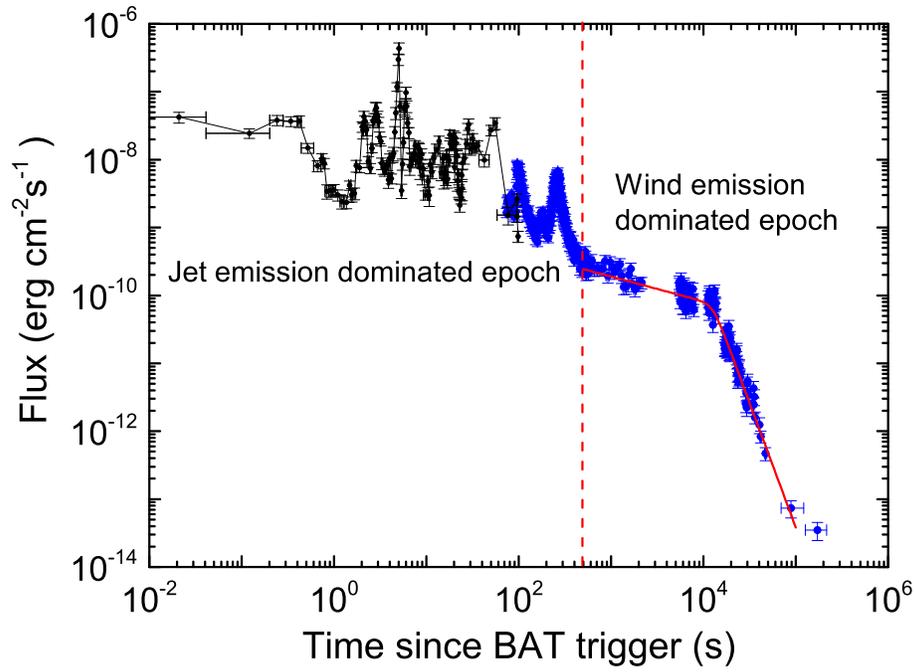}
\caption{Joint BAT+XRT lightcurve of GRB 060607A. Black dots are BAT data extrapolated to the XRT band (0.3-10 keV), and blue dots are XRT data. The solid red line is the fit with a smooth broken powerlaw function, and the vertical dashed line is separation of jet emission and wind emission epochs.}
\label{LC-060607A}
\end{figure}

Taking GRB 060607A as an example, we show its joint BAT+XRT lightcurve in Figure \ref{LC-060607A}. One can observe that it has two distinct epochs. One is dominated by prompt gamma-ray pluses and early X-ray flares. It may be dominated by the jet radiations via internal shocks or magnetic energy dissipation (e.g., M\'esz\'aros \& Rees 1993; Zhang \& Yan 2011). Another one is steady and long-lasting plateau, which may be dominated by the MD radiations (e.g., Zhang \& M\'esz\'aros 2001). Since early X-ray flares should be the same emission component as prompt gamma-rays (e.g., Burrows et al. 2005; Peng et al. 2014; Hu et al. 2014), we identify flares with $F_p/F_u>5$ as the jet prompt emission, where $F_p$ and $F_u$ are the fluxes at the peak time of the flare and the corresponding underlying power-law decay segment. We measure the duration of the jet emission from the start time of prompt gamma-ray duration ($T_{90}$) to the peak time of last flare/pulse with $F_p/F_u>5$ ($t_{\rm jet}$; e.g., Qin et al. 2013; Zhang et al. 2014). We extract the spectra of the prompt gamma-rays and the X-ray flares, and fit them with a simple power-law function $F\propto E^{-\beta_{\rm jet}}$ for deriving the fluence of the jet emission in the BAT-XRT band.

For robustly measuring the break time ($t_b$) of the shallow decay segment, the XRT data are also have good temporal coverage around $t_b$. We fit the XRT lightcurve post the jet emission epoch with a smooth broken power-law function, i.e.,
\begin{equation}
F=F_0\left [
\left (\frac{t}{t_b}\right)^{\omega\alpha_1}+\left (
\frac{t}{t_b}\right)^{\omega\alpha_2}\right]^{-1/\omega},
\end{equation}
where $\omega$ describes the sharpness of the break (taken as 3 in this analysis; Liang et al. 2007), and $\alpha_1$ and $\alpha_2$ are the decay indices before and after $t_b$, respectively. We fit the spectra of the MD emission dominated epoch with an absorbed single power-law function to derive the spectral index ($\beta_{\rm wind}$) and X-ray flux ($F_{\rm wind}$).

We finally have a sample of 117 long GRBs, as listed in Table 1. Sixty-seven GRBs out of them have redshift measure. We calculate their isotropic prompt gamma-ray/X-ray energy ($E_{\rm jet}$) with the fluences in the BAT and XRT band. The isotropic X-ray luminosity and energy released during the MD radiation epoch are calculated with $L_{\rm wind}=4\pi {D^2_L}F_{\rm wind}$ and $E_{\rm wind}=4\pi {D^2_L}F_{\rm wind} \tau$, where $\tau=t_b/(1+z)$ and $D_L$ is the luminosity distance. Our results are also reported in Table 1.
Figure \ref{Dis-L} shows the distributions of $t_{\rm jet}$ in the burst frame ($t_{\rm jet, z}$), $\tau$, and $L_w$ for the 67 GRBs with redshift measure. One can observe that the $t_{jet, z}$ values are normally in the range of $\log t_{\rm jet, z}/{\rm s}=1.68\pm 0.47$ and only three cases have a duration of several thousand seconds in the redshift-known sample. The $\tau$ distribution spreads from hundreds of seconds to several days, which can be fit with a Gaussian function $\log \tau/{\rm s}=3.61\pm 0.74$. The distribution of $L_b$ is clustered at $\log L_b/{\rm~erg\ s^{-1}}=47.75 \pm 0.79$.

Figure \ref{Dis-alpha} shows all GRBs in our sample in the $\alpha_1-\alpha_2$ plane together with the distributions of $\alpha_1$ and $\alpha_2$. The kinetic luminosity injected to the MD wind from the spin-down of a magnetar evolves as $L_{\rm k}\propto (1+t/\tau)^{-\alpha}$. The $\alpha$ value depends on energy lost behaviors during the magnetar spin-down, as mentioned in \S1. The $\alpha_2$ value thus may give information for different scenarios. One can observe that the shallow decay segments of GRBs 070110, 060602, 070616, and 060607A are almost a plateau and their $\alpha_2$ values are steeper than 3. This likely suggests that these magnetars are supra-massive and they collapse to form a black hole prior to their spin-down characteristic timescale (Troja et al. 2007; Fan et al. 2013; Du et al. 2016; Chen et al. 2017). The sharp drop of the X-ray emission may indicate that the kinetic luminosity injected to the MD wind is rapidly turned off. The curvature effect to the high latitude emission after the cease of the emission results in a temporal evolution feature of the observed flux as $F\propto t^{-(2+\beta_{\rm wind})}$ (e.g. Dermer 2004). An extremely steep decay slope would be due to the zero time effect (Liang et al. 2006). The $\alpha_2$ values of 12 GRBs are in the range of $2<\alpha_2<3$. They are also roughly consistent with the curvature effect or due to the energy lost of the magnetar spin-down are
dominated by the MD wind. The $\alpha_2$ values of most GRBs are of $1<\alpha_2<2$. They are consistent with the generic scenario that the spin-down energy release via both the GW and MD radiations for a stable magnetar.

\section{Central Engine Properties and Connections with Jet emission}
We estimate the initial spin period ($P_0$), the surface polar cap magnetic field strength ($B_p$) of the magnetars of the GRBs in our sample with the observed $L_b$ and $\tau$. Following Zhang \& M\'{e}sz\'{a}ros (2001), we have
\begin{eqnarray}
L_{\rm k} = 1.0 \times 10^{49}~{\rm erg~s^{-1}} (B_{\rm p,15}^2 P_{0,-3}^{-4} R_6^6),\\
\tau = 2.05 \times 10^3~{\rm s}~ (I_{45} B_{\rm p,15}^{-2} P_{0,-3}^2 R_6^{-6}),
\label{L0}
\end{eqnarray}
where $I$ is the inertia moment, $R$ is the radius of the magnetar, and $Q_n=Q/10^{n}$ in cgs units. Based on Eqs.(2) and (3), one has
\begin{eqnarray}
B_{\rm p,15} = 2.05(I_{45} R_6^{-3} L_{k,49}^{-1/2} \tau_{3}^{-1})~\rm G,\\
P_{0,-3} = 1.42(I_{45}^{1/2} L_{k,49}^{-1/2} \tau_{3}^{-1/2})~\rm s.
\label{Bp-tau}
\end{eqnarray}

We take the radiation efficiency of the MD wind as 0.3 (Du et al. 2016), i.e., $L_{\rm b}= 0.3L_{\rm k}$, $I=10^{45}$ g cm$^{2}$, and $R=10^{6}$ cm for deriving the $P_0$ and $B_p$ values. Our results are presented in Table 2. The derived $P_0$ and $B_p$ values are in the ranges of $P_0\in(0.6, 144.1)$ ms and $B_p\in(0.26, 22.40)\times 10^{15}$ G. They are comparable to that reported by L\"{u} \& Zhang (2014). We do not find any correlation of these parameters with $\alpha_2$ and $\tau$, as shown in Figure \ref{Magnetar}. $P_0$ and $B_p$ as a function of $E_{\rm jet}$ are shown in Figure \ref{Magnetar-jet}. Interestingly, $P_0$ is tightly correlated with $E_{\rm jet}$, with a Spearman linear correlation coefficient $r=-0.83$ and a chance probability $p<10^{-4}$. Our linear fit with the least square regression algorithm yields
\begin{equation}
\log P_0 =(19.39\pm 1.62)-(0.36\pm 0.03)\log E_{\rm jet}.
\end{equation}
This relation may suggest that a magnetar with faster rotation speed can power a more energetic jet. No statistical correlation between $B$ and $E_{\rm jet}$ can be claimed. These results suggest that $P_0$ would be essential for jet production.

\section{Relation of Energy Releases between the Jet and Wind}
Energy partition between the jet and MD wind is also of our interest. Figure \ref{Jet-wind} shows $E_{\rm wind}$ as a function of $E_{\rm jet}$. One can observe that the two quantities are correlated. The Spearman correlation analysis yields a linear correlation coefficient $r=-0.85$ and chance probability $p<10^{-4}$. Our linear fit with the least square regression algorithm gives \begin{equation}
\log E_{\rm wind}/{\rm erg}=(3.11\pm 3.72)+(0.91\pm 0.07)\times \log E_{\rm jet}.
\end{equation}
This correlation may imply that the energy partition between the jet and MD wind is quasi-universal among these GRBs. Since $E_{\rm wind}$ is roughly proportional to $E_{\rm jet}$, we measure the energy partition with a ratio of $R\equiv E_{\rm wind}/E_{\rm jet}$. The distributions of $E_{\rm wind}$, $E_{\rm jet}$, and $R$ together with the Gaussian fits are shown in Figure \ref{R}. The Gaussian fits yield $\log R=(-1.62\pm0.50)$, $\log E_{\rm wind}/{\rm erg}=50.96\pm0.22$, and $\log E_{\rm jet}/{\rm erg} =52.54\pm0.43$. Typically, $E_{\rm jet}$ is about two orders of magnitude larger than $E_{\rm wind}$ and the derived typical $R$ value is $R=0.03$. Note that GRB jets are highly collimated and the MD wind may be quasi-isotropic. The true jet energy is $E_{\rm jet}=E_{\rm jet}(1-\cos\theta_{\rm jet})\approx E_{\rm jet}\theta_{\rm jet}^2/2$, where $\theta_{\rm jet}$ is the jet opening angle in unit of rad. By making geometrical correction for $\log E_{\rm jet}$ with a typical jet opening angle of $10^{o}$ (e.g., Frail et al. 2001; Liang \& Zhang 2005), we have $R\sim 2$. This hints that the energies of the jet and MD wind would be comparable.

\section{Misaligned Magnetar jet: Origin of Some Ultra-long GRBs?}
As mentioned in \S1, GRBs 101225A and 170714A show up as ultra-long GRBs. Their gamma-ray/X-ray emission is long-lasting and steady (Campana et al. 2011; Th\"{o}ne et al. 2011). The emission of GRB 101225A was detected 80 seconds prior to the BAT trigger time ($T_0$) and lasted up to $T_0+1672$ seconds (Cummings et al. 2010). We set the zero time of this event at $T_0-80$ seconds. Its joint BAT-XRT lightcurve features as a plateau with significant flickers/flares and a sharp drop at $T_0+2\times 10^4$ seconds (Palmer et al. 2010). The global lightcurve can be fit by a broken power-law with index changing from $\alpha_1=0.12\pm0.06$ to $\alpha_2=6.46\pm 0.39$ broken at $T_0+2\times 10^4$ s. The mask-weighted BAT light curve of GRB 170714A shows continuous weak emission starting at about $T_0-70$ seconds. We therefore set its zero time as $T_0-70$ seconds. Its BAT lightcurve is steady with a power-law index $\alpha_1=0.20\pm 0.12$ until the end of the BAT event data ($\sim 960$ seconds), and its joint BAT-XRT lightcurve illustrates a clear drop with a slope of $\alpha_2=4.70\pm 0.13$.

The joint BAT+XRT lightcurves of GRBs 101225A and 170714A are similar to the X-ray plateaus observed in some typical GRBs, such as GRBs 060607A, 070110, 100814A, and 151027A, as shown in Figure \ref{ultra-long}. We add GRBs 101225A and 170714A to Figures \ref{Dis-alpha}. One can observe that they are similar to GRBs 060602A, 060607A, 070110, and 070616. We compare their lightcurves with the shallow-decay segments of some typical GRBs in Figure \ref{ultra-long}. They are apparently consistent. Non-detection of their jet emission may be due to off-axis observations to their jets. We thus estimate the $P_0$ and $B_p$ values of these magnetars.  By adding them in Figures \ref{Magnetar}-\ref{Jet-wind} in comparison with the typical GRBs in our sample, it is found that they are not distinct from the other GRBs. We estimate the energy releases of their MD winds and obtain $E_{\rm wind}=7.86\times 10^{51}$ erg of GRB 101225A and $E_{\rm wind}=9.19\times 10^{51}$ erg of GRB 170714A. Assuming that they share the same energy partition of the jet and wind as typical long GRBs, i.e., $R=0.03$, we infer their $E_{\rm jet}$ values as $\sim2.62\times 10^{53}$ erg and $\sim3.06\times 10^{53}$ erg, respectively. Therefore, their jet prompt emission may be potentially very bright and may be located at the high $E_{\rm jet}$ end of the $E_{\rm wind}-E_{\rm jet}$ relation, as shown in Figure \ref{Jet-wind}.

\section{Discussion}
\subsection{Observational Biases}
Our analysis presents a tight $P_0-E_{\rm jet}$ relation based on a sample of those GRBs that have a shallow decay segment in their early XRT lightcurves. This sample suffers significant observational biases since observations with BAT and XRT depend on their flux thresholds (e.g., Bulter et al. 2009).  Butler et al. (2009) presented a general approach for evaluating the impact of detector threshold truncation to apparent correlations of GRBs. The determination of the true source frame relation requires knowledge of the GRB rate density and luminosity function (LF) to impute the missing data (Dainotti et al. 2015). We here follow the same approach as Dainotti et al. (2015) to discuss whether the observational biases can result in the $P_0-E_{\rm jet}$ relation.

\begin{figure}
\includegraphics[angle=0,scale=0.3]{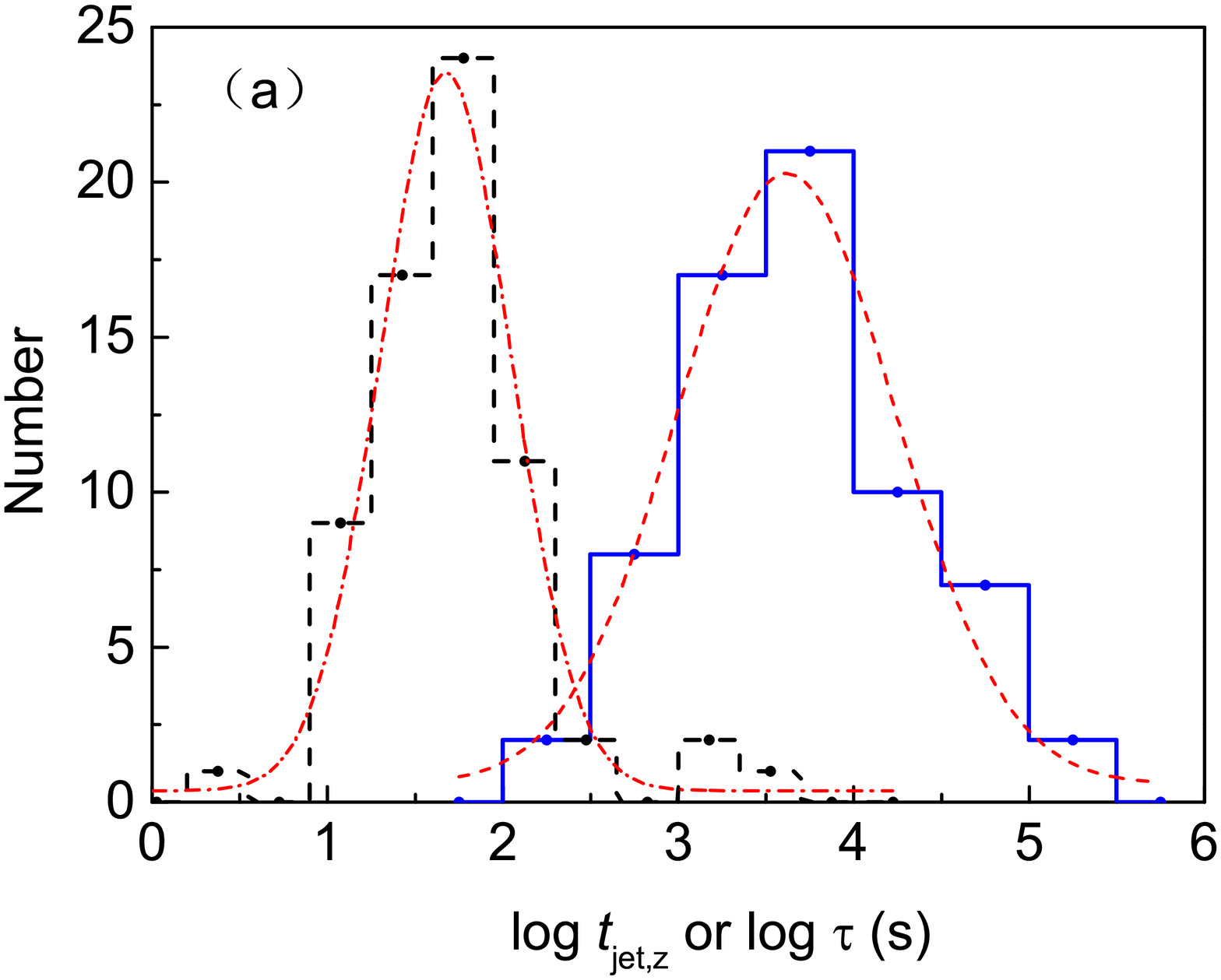}
\includegraphics[angle=0,scale=0.3]{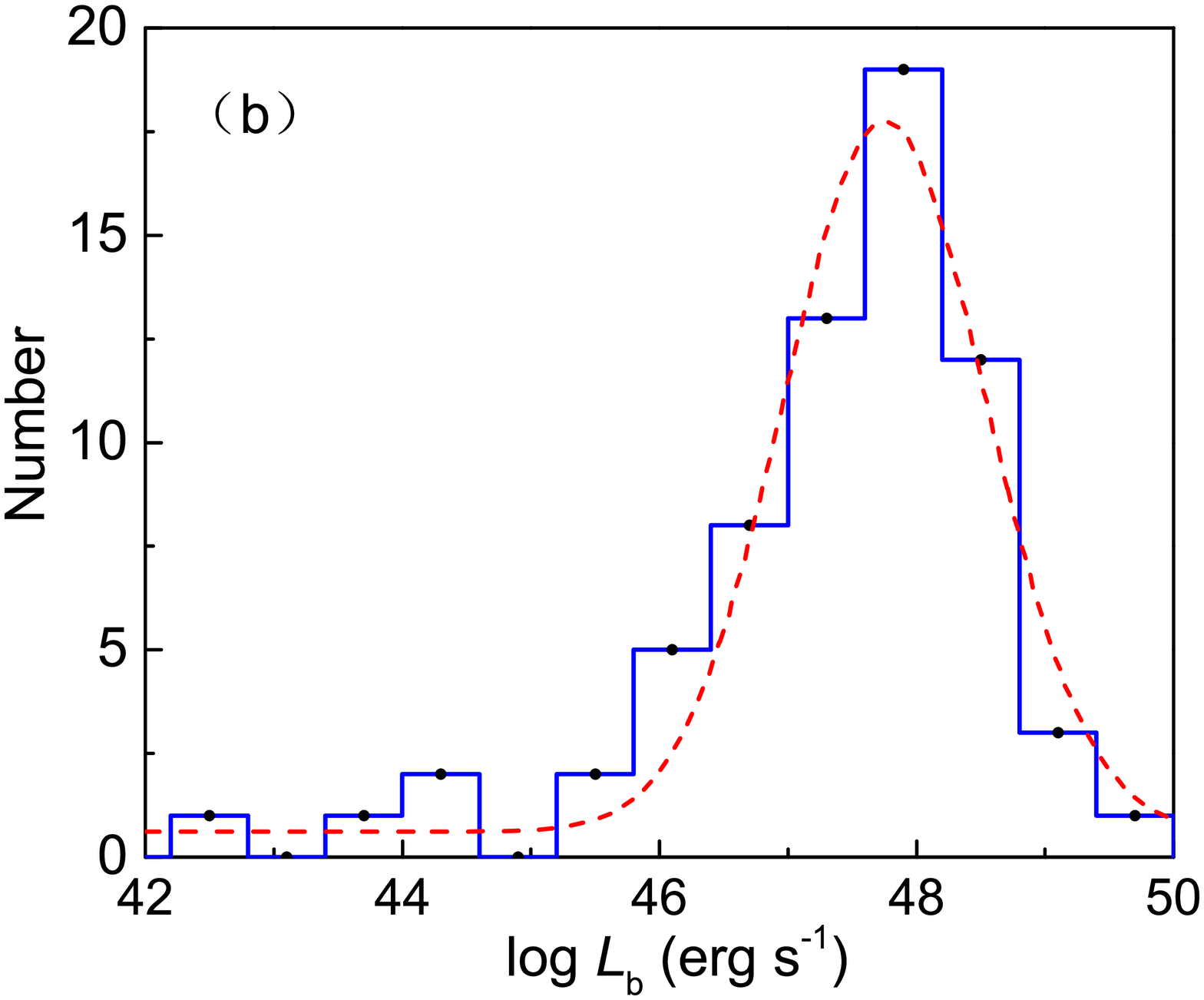}
\caption{Distributions of jet emission duration ($t_{\rm jet}$) and the duration of the wind emission measured with the break time ($t_b$) in our fits in the rest frame ({\em left panel}) as well as plateau luminosity $L_b$ ({\em right panel}). The dashed lines are the best Gaussian fits.}
\label{Dis-L}
\end{figure}

\begin{figure}
\includegraphics[angle=0,scale=0.5]{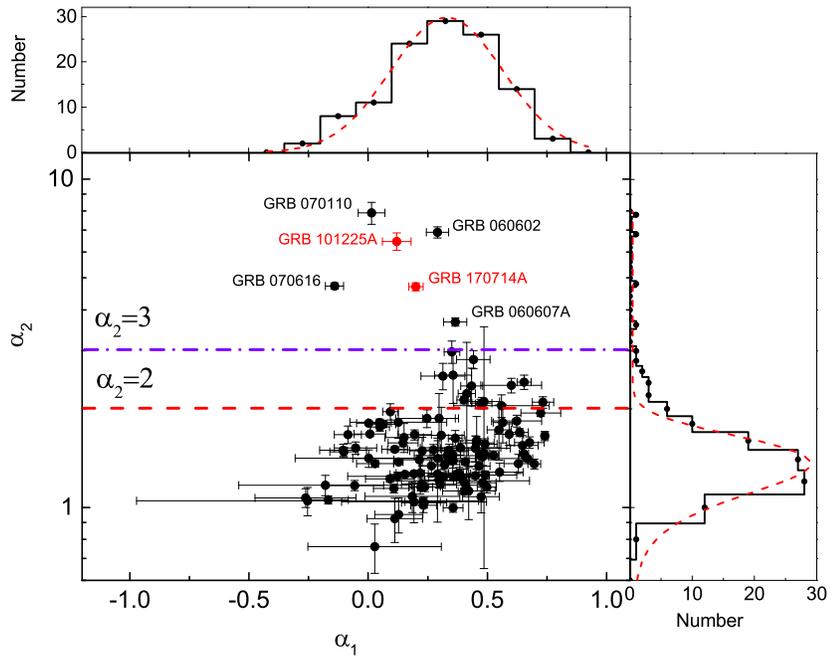}
\caption{ One- and two-dimensional distributions of $\alpha_{1}$ and $\alpha_{2}$ in our sample.
The red stars are ultra-long GRBs 101225A and 170714A. Two horizontal dashed lines
correspond to $\alpha_2=2$ and $\alpha_2=3$, respectively. The Gaussian fits to the distributions are also shown with dashed lines.}
\label{Dis-alpha}
\end{figure}

\begin{figure}
\includegraphics[angle=0,scale=0.3]{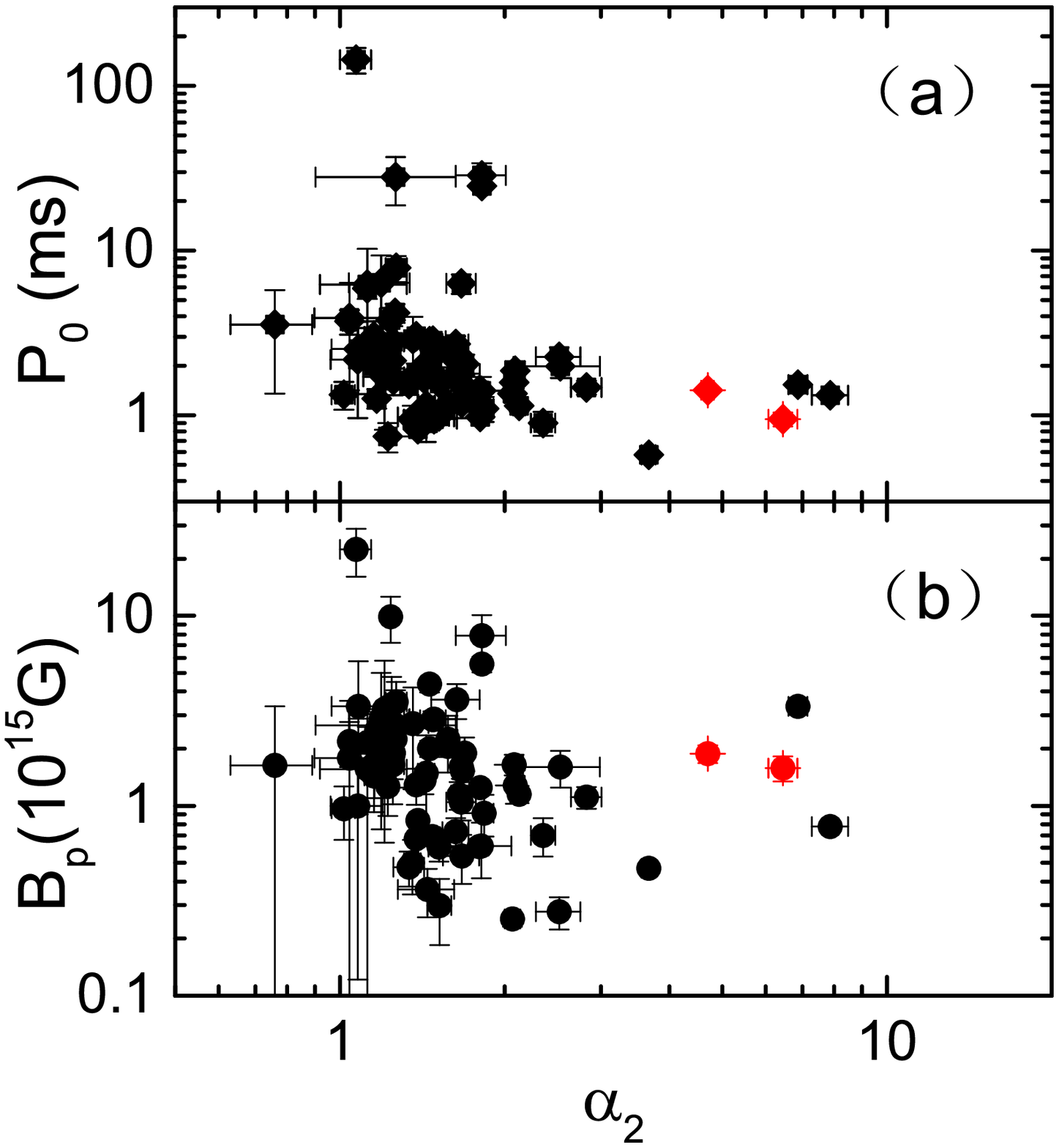}
\includegraphics[angle=0,scale=0.3]{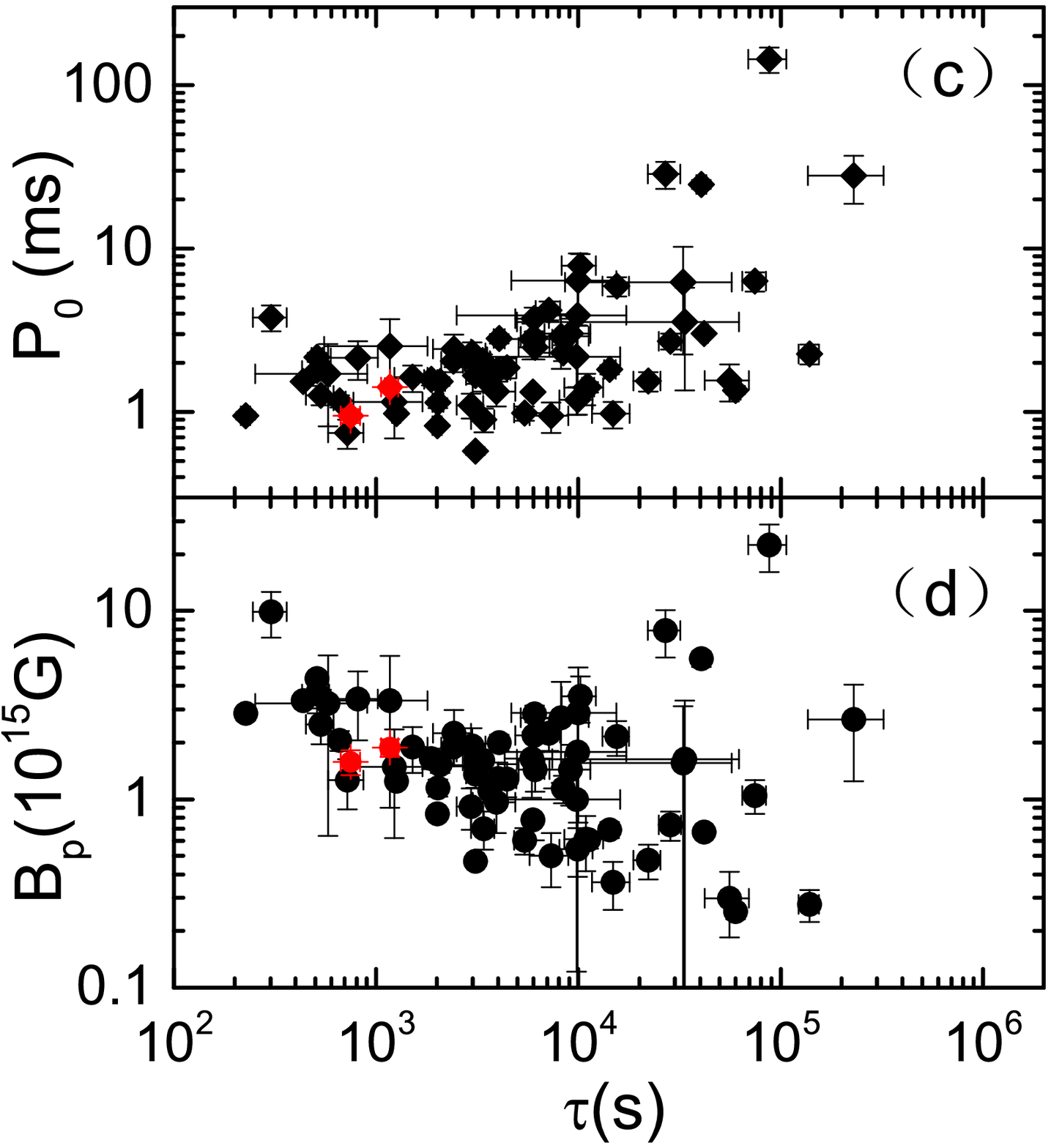}
\caption{$P_0$ and $B_p$ as a function of $\alpha_2$ and $\tau$, respectively. The dots and squares on behalf of $B_p$ and $P_0$.
The red stars are GRBs 101225A and 170714A.}
\label{Magnetar}
\end{figure}

\begin{figure}
\includegraphics[angle=0,scale=0.3]{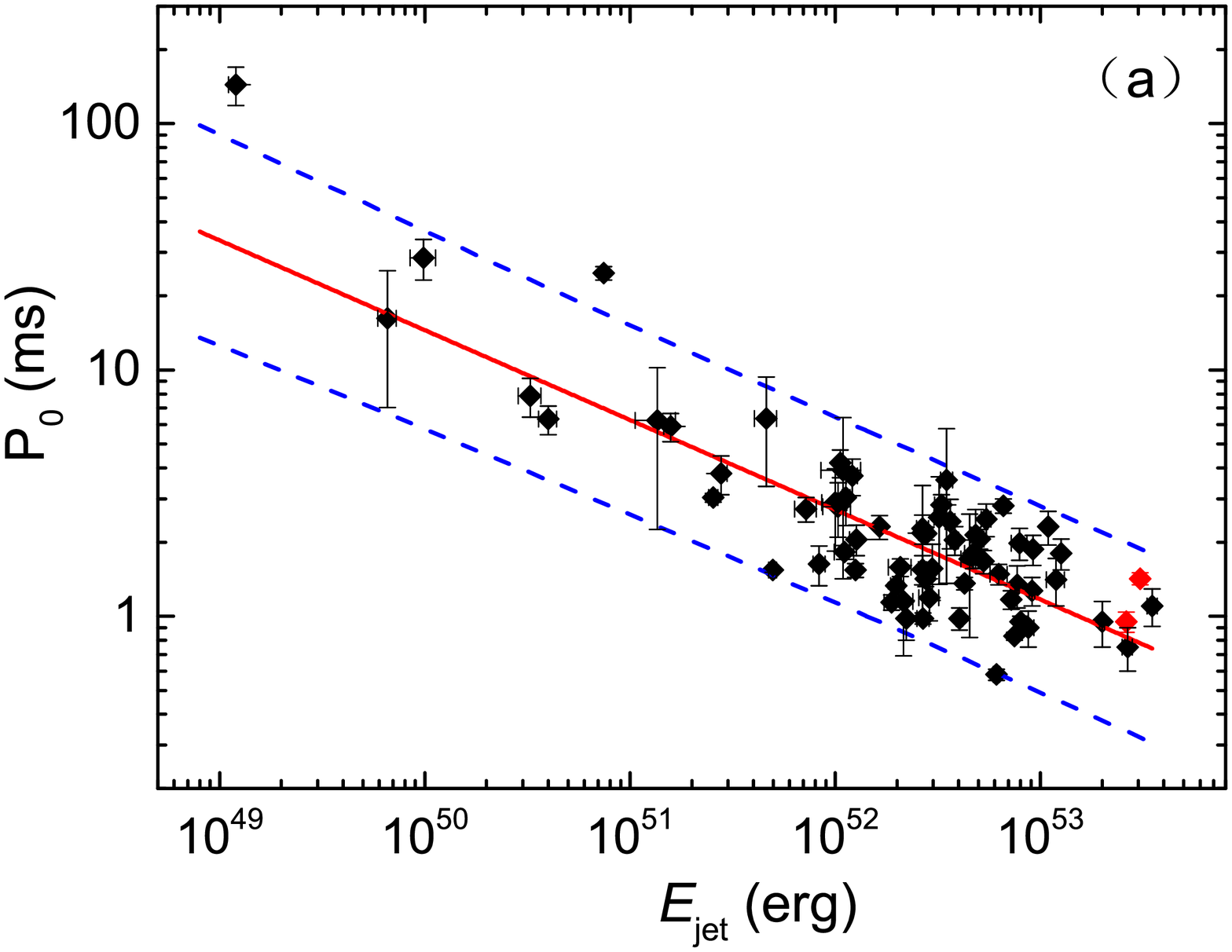}
\includegraphics[angle=0,scale=0.3]{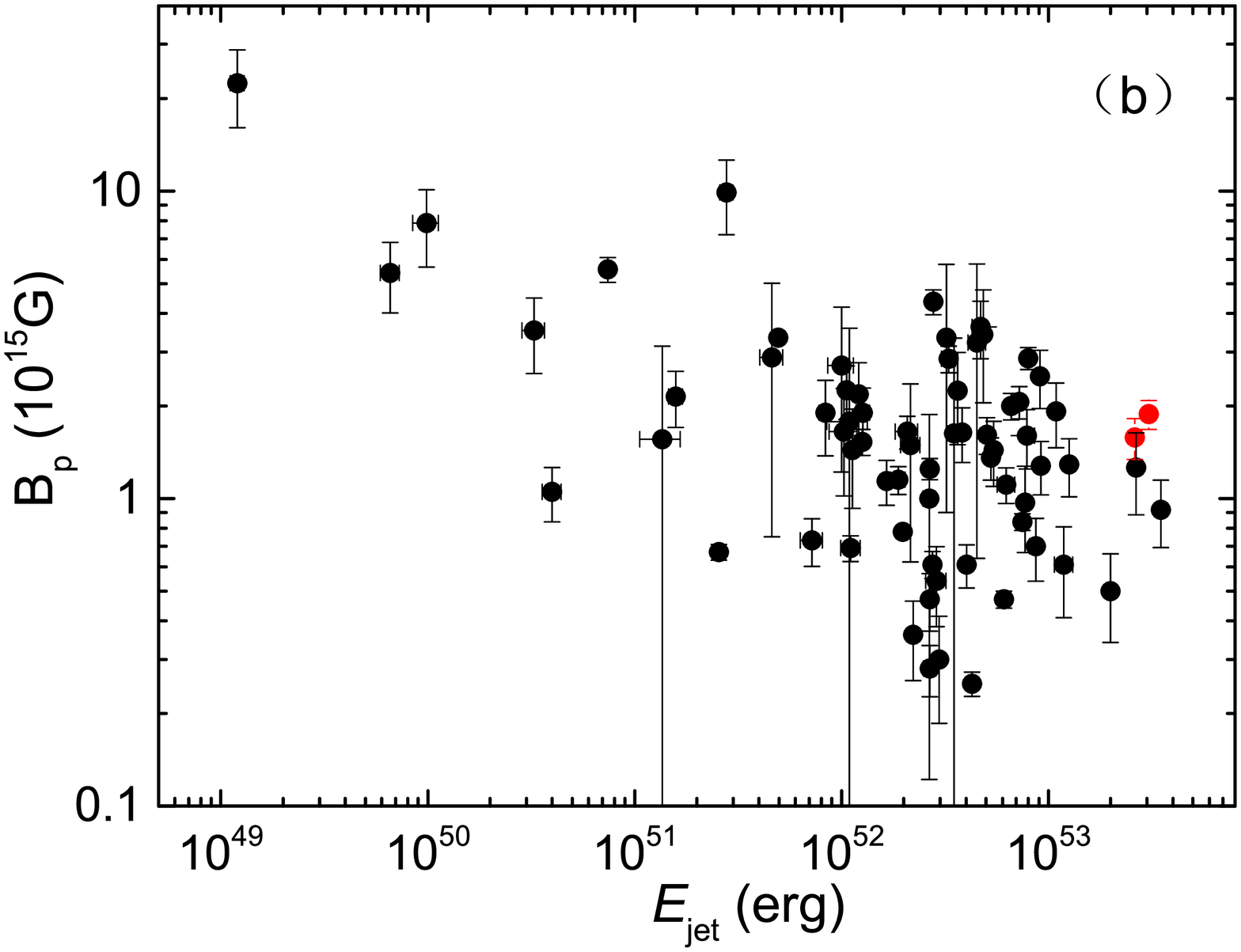}
\caption{$P_0$ and $B_p$ as a function of $E_{\rm jet, iso }$. The solid lines are the least square linear fits, and two dashed lines are 95\% confidence level of the fits. The dots and squares on behalf of $B_p$ and $P_0$. The red dots are for GRBs 101225A and 170714A. The gray dots are the result of considering the threshold.}
\label{Magnetar-jet}
\end{figure}

\begin{figure}
\includegraphics[angle=0,scale=0.5]{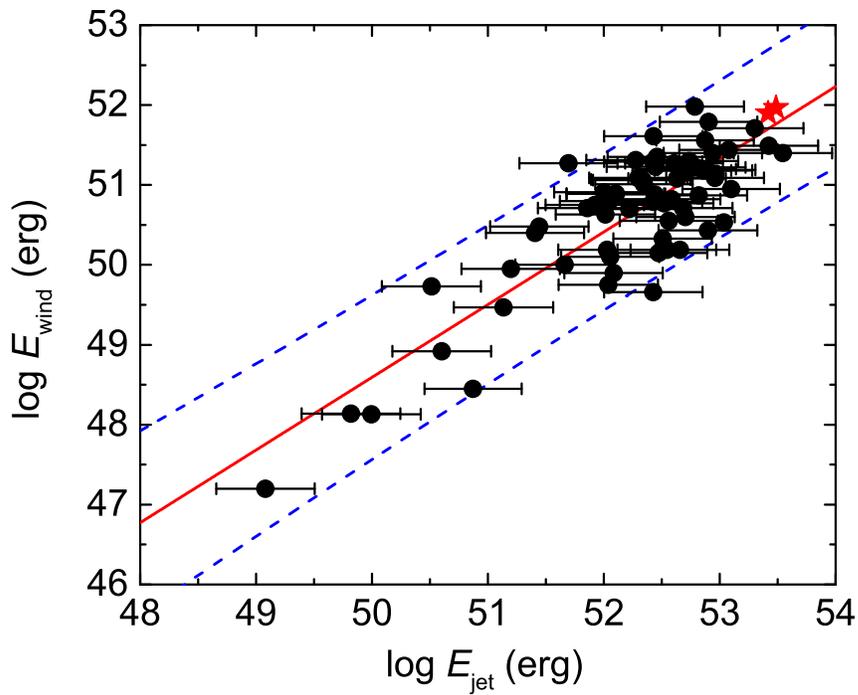}
\caption{Correlation between $E_{\rm jet}$ and
$E_{\rm wind}$. The solid line and dashed lines are the least square fit and the $95\%$ confidence level of the fits, respectively. Two red stars are GRBs 101225A and 170714A.}
\label{Jet-wind}
\end{figure}

\begin{figure}
\includegraphics[angle=0,scale=0.3]{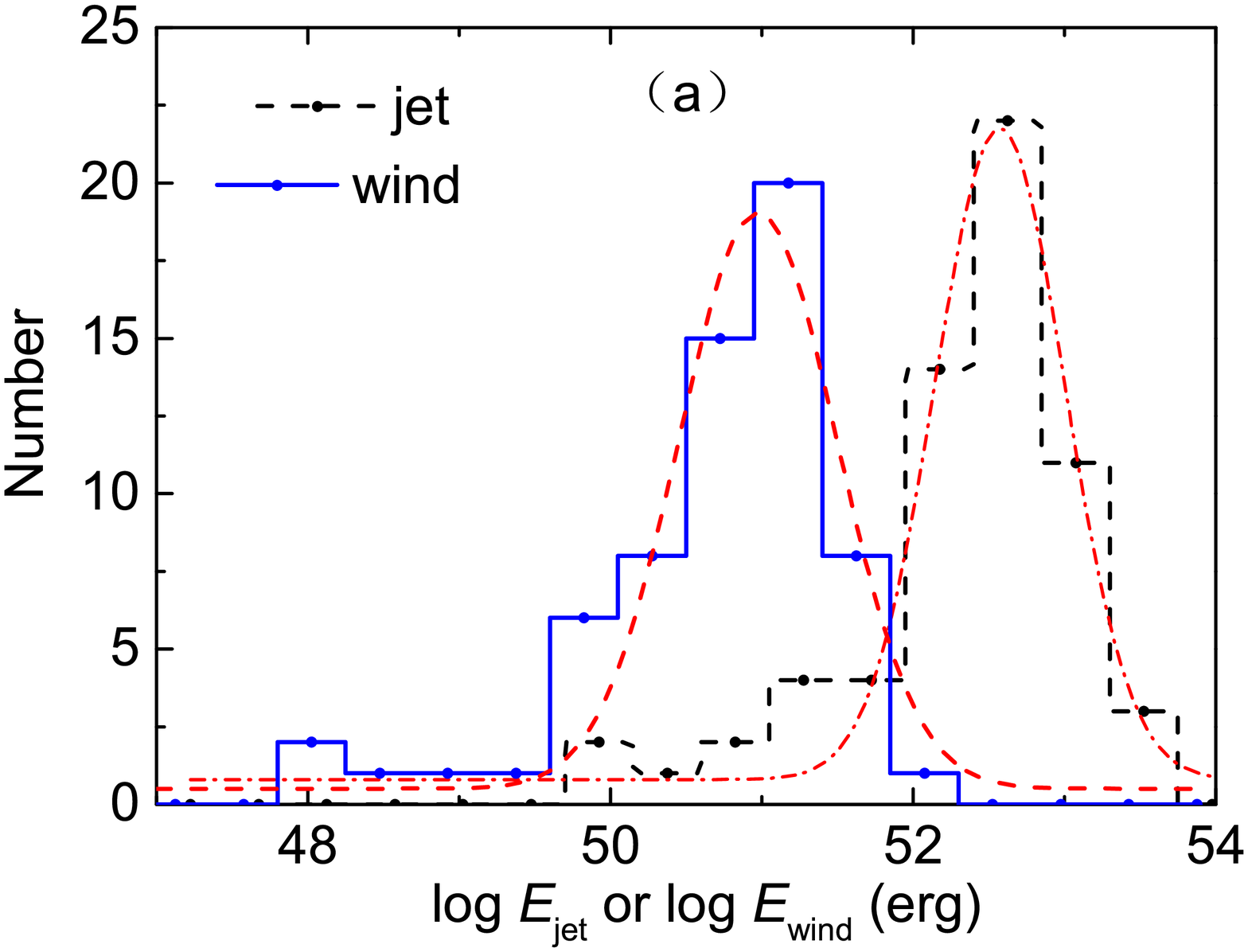}
\includegraphics[angle=0,scale=0.3]{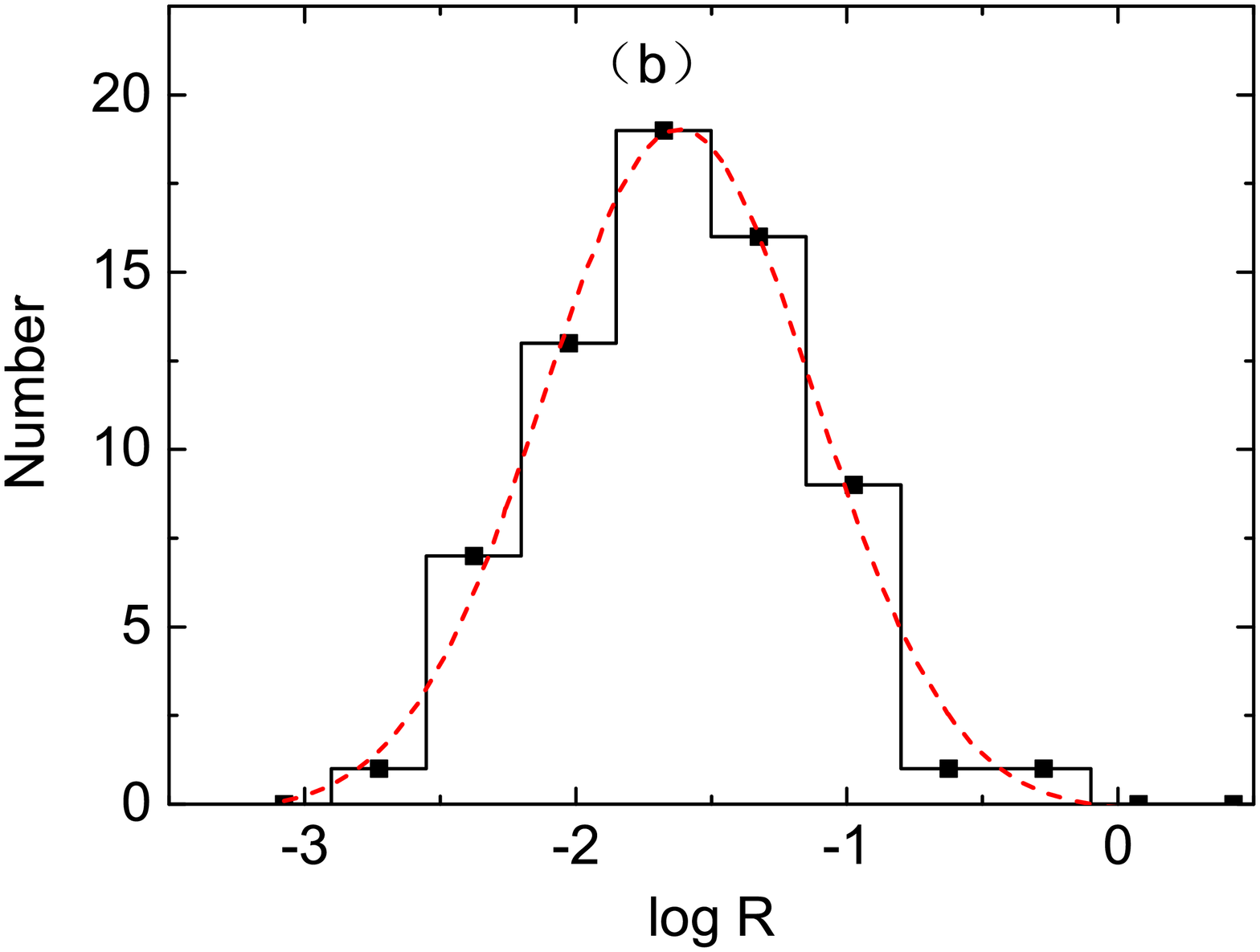}
\caption{{\em Panel (a)---} Distributions of isotropic prompt gamma-ray/X-ray energy ($E_{\rm jet}$) and X-ray energy release of the magnetar MD wind ($E_{\rm wind}$) for the GRBs in our sample. {\em Panel (b)---} Distribution of the energy partition ratio $R$ for the GRBs in our sample. Dashed lines are the best Gaussian fits to the distributions.}
\label{R}
\end{figure}

\begin{figure}
\includegraphics[angle=0,scale=0.5]{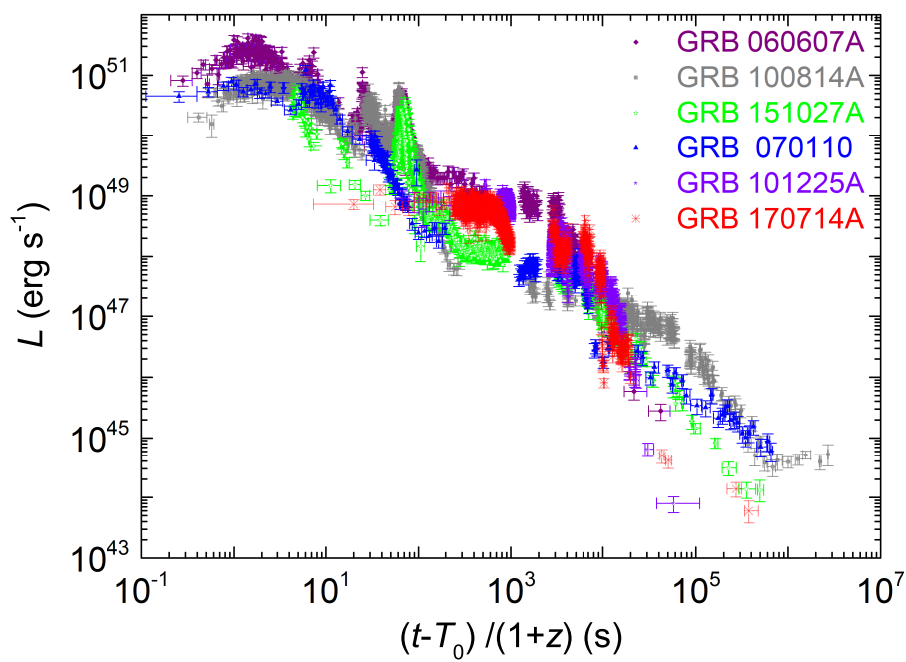}
\caption{Joint BAT+XRT lightcurves of ultra-long GRBs 101225A and 170714A in comparison with some typical-long GRBs that have a clear X-ray plateau detected. Note that the zero time ($T_0$) of these lightcurves are shifted to prior the BAT trigger time since the signals were clearly detected prior the BAT trigger (e.g. Hu et al. 2014).}
\label{ultra-long}
\end{figure}

\begin{figure}
\includegraphics[angle=0,scale=0.3]{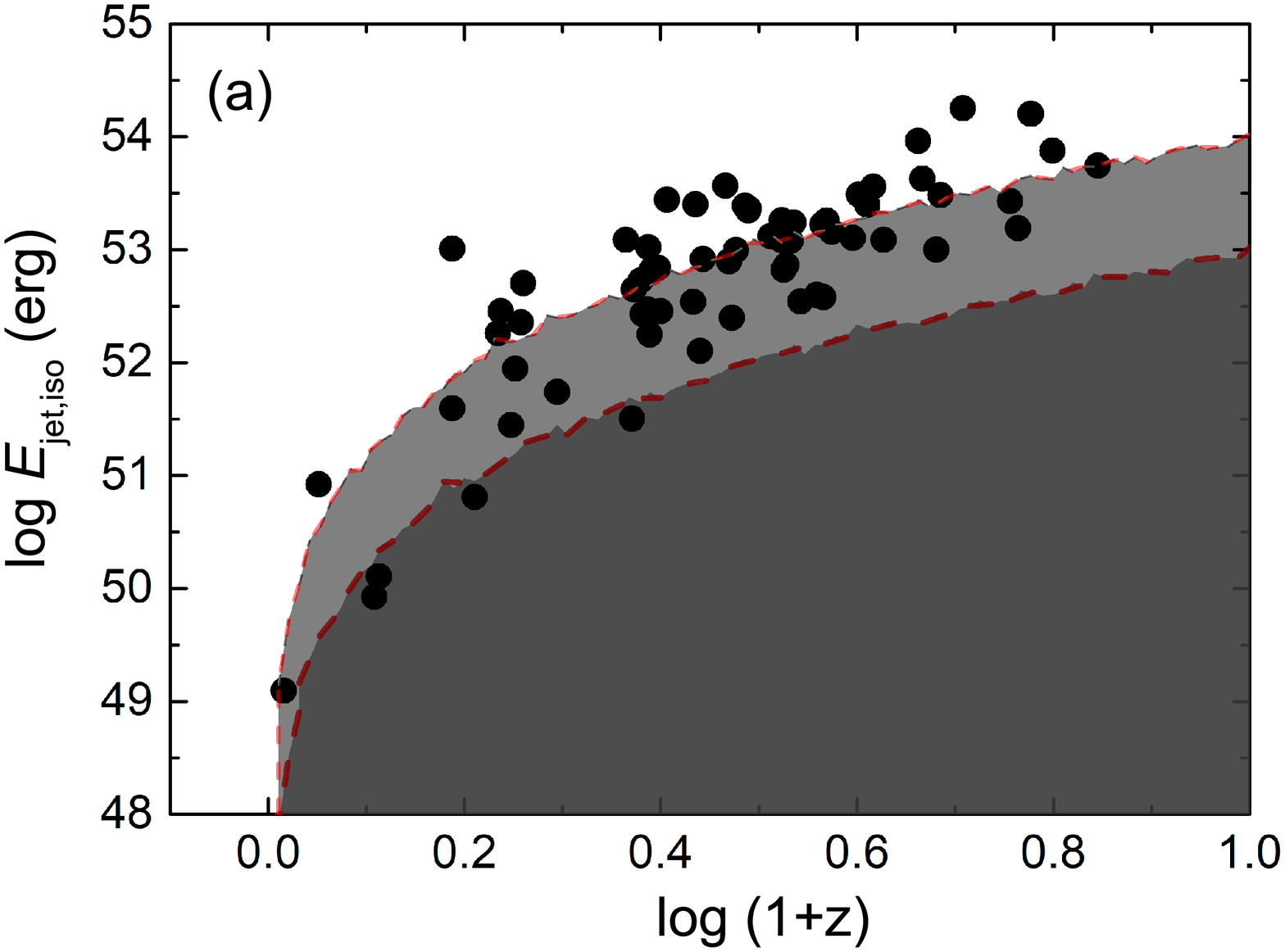}
\includegraphics[angle=0,scale=0.3]{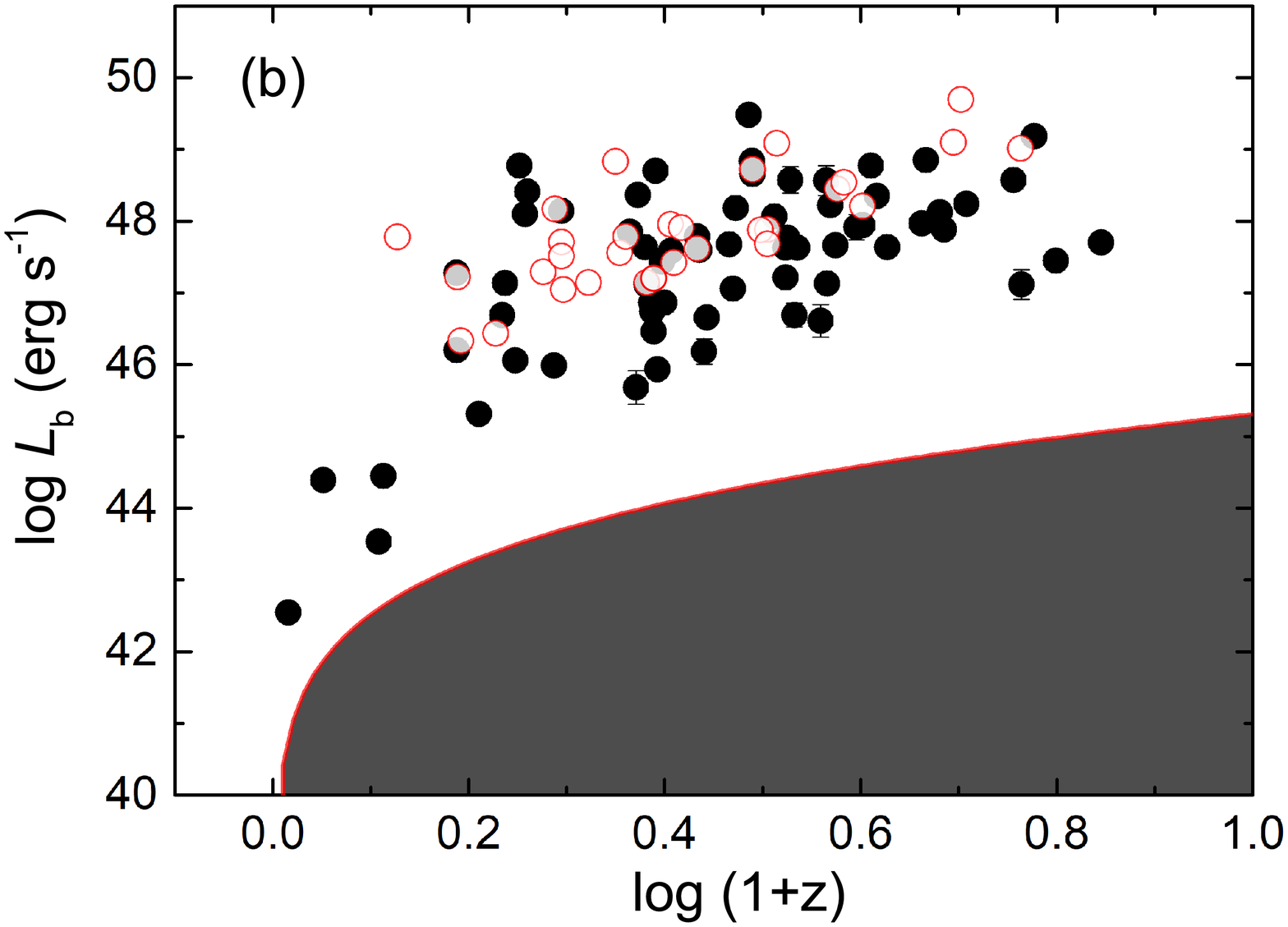}
\caption{{\em Panel (a)---} GRB energy thresholds of {\em Swift}/BAT in the count rate mode for directly on-axis GRBs with a flux limit of $F^{\rm th, on}_{\rm BAT}=1.0\times10^{-8}$ erg cm$^{-2}$ s$^{-1}$ and for extremely off-axis GRBs with a flux limit of $F^{\rm th, off}_{\rm BAT}=1.0\times10^{-7}$ erg cm$^{-2}$ s$^{-1}$ (Lien et al. 2014). The duration of the GRB emission is bootstrapped from a log-normal distribution of $\log t_{\rm jet, s}/{\rm s}=1.68\pm 0.47$. The GRBs in our sample are shown as dots. GRBs with a flux being lower than the flux limits for on-axis GRBs may be triggered in the image mode. {\em panel (b)---} Luminosity threshold of {\em Swift}/XRT for a flux limit of $F^{\rm th}_{\rm XRT}=2.0\times10^{-14}$ erg cm$^{-2}$ s$^{^-1}$. The data of the X-ray plateaus in our sample are shown with solid dots, and data of the X-ray afterglow luminosity at $T_0+3600$ seconds for the GRBs with a single power-law XRT lightcurve are shown with opened circles.}
\label{threshold}
\end{figure}

The lowest flux truncation of BAT is $F^{\rm th, on}_{\rm BAT}=1\times10^{-8}$ erg cm$^{-2}$ s$^{-1}$ for GRBs with an incident angle of zero (perfectly on-axis GRBs). However, most GRBs occur at larger incident angles (off-axis GRBs). For extremely off-axis events, BAT trigger threshold could be lowered down to  $F^{\rm th, on}_{\rm BAT}=1\times10^{-7}$ erg cm$^{-2}$ s$^{-1}$ for GRBs with an incident angle of $55^{o}$ (Lien et al. 2014). The flux truncation of XRT is $F^{\rm th}_{\rm XRT}=2\times 10^{-14}$ erg cm$^{-2}$ s$^{-1}$. Accordingly, we have $E^{\rm th}_{\rm jet}=4\pi D^{2}_L F^{\rm th}_{\rm BAT}t_{j,z}$ and $L^{\rm th}_b=4\pi D^{2}_L F^{\rm th}_{\rm XRT}$, where $t_{\rm jet, z}$ is generated from the $t_{\rm jet, z}$ distribution of our sample as shown in Figure \ref{Dis-L}, i.e., $\log t_{\rm jet, z}/{\rm s}=1.68\pm 0.47$, via a bootstrap algorithm. Figure \ref{threshold} (a) shows the BAT detection threshold and the GRBs in our sample. Note that the trigger probability of a GRB with a flux level close to the threshold in the count rate trigger mode is low (e.g., Butler et al. 2009; Qin et al. 2010; Coward et al. 2013; Dainotti et al. 2015). In addition, the lowest flux truncation of $F^{\rm th, on}_{\rm BAT}=1\times10^{-8}$ $erg$ cm$^{-2}$ s$^{-1}$ is for perfectly on-axis GRBs, and most GRBs occur at larger incident angles. These effects lead to most GRBs in our sample are significantly over the lowest threshold threshold of BAT. A small fraction of GRBs are close or even below the detection threshold. These GRBs are usually very long and triggered by the image mode (Sakamoto et al. 2009). Figure \ref{threshold}(b) shows the XRT detection threshold and the X-ray plateau data of GRBs in our sample. The XRT data are also much higher than the XRT threshold line since the XRT data are obtained by the follow-up observations to the BAT trigger, but not from an independent blind survey with the XRT sensitivity.

\begin{figure}
\includegraphics[angle=0,scale=0.5]{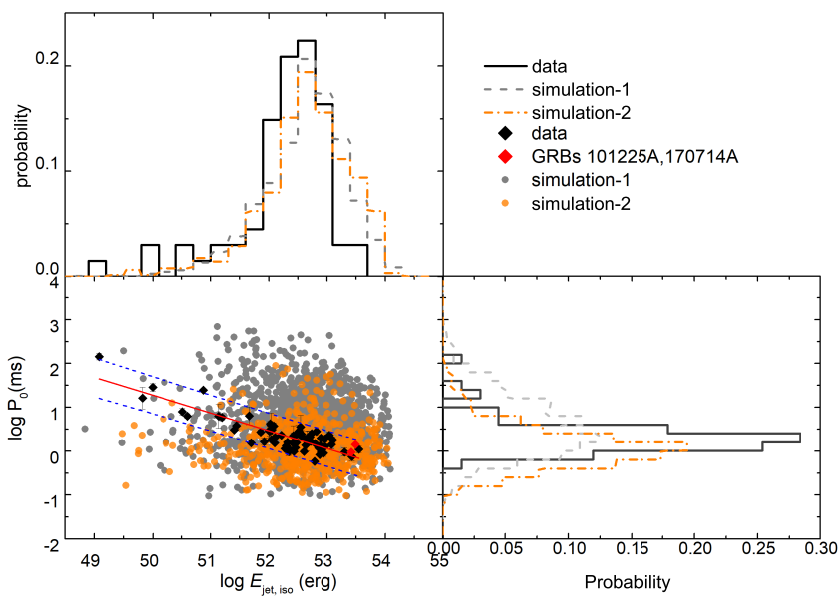}
\caption{Comparison between the observed and simulated GRB samples in the $\log P_0-\log E_{\rm jet}$ plane and in the one-dimensional $\log P_0$ and $\log E_{\rm jet}$ distributions. The grey dots (``simulations-1") are for simulations by considering only the BAT and XRT flux limits, and the yellow dots (``simulations-2") are for simulations further consideration of the jet afterglow contaminations. }
\label{simulation}
\end{figure}

To evaluate the instrumental biases, we make a Monte Carlo simulation analysis with an approach as Qin et al. (2010). We outline the procedure of our simulations as following.
\begin{itemize}
\item We assume that the GRB rate as a function of redshift follows the star formation rate (SFR) and adopt an SFR parameterized form reported by  Hopkins \& Beacom (2006). The local GRB rate is taken as 1.12 Gpc yr$^{-1}$ (e.g., Liang et al. 2007).
\item The LFs of both the GRBs and dipole wind emission are taken as $\Phi(L)=\Phi_{0}[(L/L_{c})^{\alpha_1}+(L/L_{c})^{\alpha_2}]^{-1}$, where $\Phi_{0}$ is the normalization parameter and $\alpha_1$ and $\alpha_2$ are the power-law indices breaking at $L_c$. We take the distribution of the jet emission epoch as a log-normal distribution of $\log t_{jet,z}/\rm s=1.68\pm 0.47$ (1$\sigma$). The $E_{\rm jet}$ of a mock GRB is calculated with $E_{\rm jet}=L_{\rm jet}\times t_{jet,z}$.  The flux limit of BAT is randomly picked up in the range between  $F^{\rm th, off}_{\rm BAT}$ and $F^{\rm th, on}_{\rm BAT}$ (Lien et al. 2014).
\item We constrain the parameters of the GRB LF by measuring the consistency between the $E_{\rm jet}$ distributions of the simulated sample and the observed sample with the Kolmogorov¨CSmirnov (K-S) algorithm by adopting a p-value of the K-S test as $P_{\rm K-S}>10^{-4}$, as shown in Figure \ref{simulation}. We get $\alpha^{\rm jet}_1=0.65$, $\alpha^{\rm jet}_2=2.3$, $L^{\rm jet}_{c}=1.25\times10^{51}$ erg s$^{-1}$ . The LF parameters of the MD wind emission are taken as $\alpha^{\rm wind}_1=0.8$, $\alpha^{\rm wind}_2=1.8$, $L^{\rm wind}_{c}=1.0\times10^{48}$ erg s$^{-1}$ (Xie et al. 2019, in preparation). The distribution of the spin-down characteristic timescale is log-normal, i.e., $\log \tau/\rm s=3.61\pm 0.74$, as shown in Figure \ref{Dis-L}.
\item We generate a set of $\{z, L_{\rm jet}, t_{\rm jet, z}, E_{\rm jet}, L_{\rm wind}, \tau, P_0\}$ for a mock GRB based on the SFR, the LFs of the GRB jet and dipole radiation wind, the distributions of $t_{\rm jet,z}$ and $\tau$, where $P_0$ is calculated with Eq. \ref{Bp-tau}. We pick up a mock GRB that its prompt gamma-ray emission and the MD wind emission are detectable with BAT and XRT, respectively.
\end{itemize}
We simulate a mock sample of 1500 GRBs. Figure \ref{simulation} shows $P_0$ as a function of $E_{\rm jet}$ for the mock GRB sample in comparison with the observed GRB sample. One can observe that the instrumental selection effect only cannot explain the observed $P_0-E_{\rm jet}$ correlation.

Detection or not of a shallow decay segment in the XRT lightcurve also depends on the fluxes competition between the MD wind emission and the GRB jet afterglow emission. Besides the canonical XRT lightcurves, the XRT lightcurves of some GRBs illustrate as a single power-law (SPL) function from tens or hundreds seconds to $\sim 10^5$ seconds post the GRB triggers, such as GRB 061007 (Liang et al. 2009), GRB130427A (Maselli et al. 2014), and GRB160625B (Troja et al. 2017). They are well interpreted as the afterglows of the GRB jets (e.g., Liang et al. 2009). They are apparently different from the canonical one. Two possibilities may address this difference. One is that the central engines of GRBs with a SPL XRT lightcurve are not a magnetar, but a black hole. In this scenario, the SPL XRT lightcurves are attributed to X-ray afterglows from external shocks of the jets without an extra emission component as that provides by the MD wind. The other one is that the central engines of the GRBs with a SPL XRT lightcurve are still a magnetar, but their MD wind emission is much lower than the jet X-ray afterglows. The MD wind emission thus may be fully buried under the bright X-ray afterglows. The luminosity of the SPL lightcurves at the early stage, such as $t-T_0<10^2$ seconds, is usually brighter than the canonical ones (Liang et al. 2009). This is really true for some energetic GRBs, such as GRB 061007 (Liang et al. 2009), GRB130427A (Maselli et al. 2014), and GRB160625B (Troja et al. 2017). Since the X-ray afterglow flux usually decays as $t^{-1}$ but the injected MD wind kinetic luminosity evolves as $L_k\propto (1+t/\tau)^{-2}$, those MD wind emission with a short $\tau$ could be rapidly decay and covered by the afterglows. If $L_k$ and $\tau$ of the GRBs with a bright SPL XRT lightcurve is weak and short, energy partition into the MD wind should be low in comparison to that into the jet. This is not consistent with the correlation of Eq. (\ref{Jet-wind}), deviating the quasi-universal energy partition between the jet and MD wind. This challenges the results in this analysis.

\begin{figure}
\includegraphics[angle=0,scale=0.5]{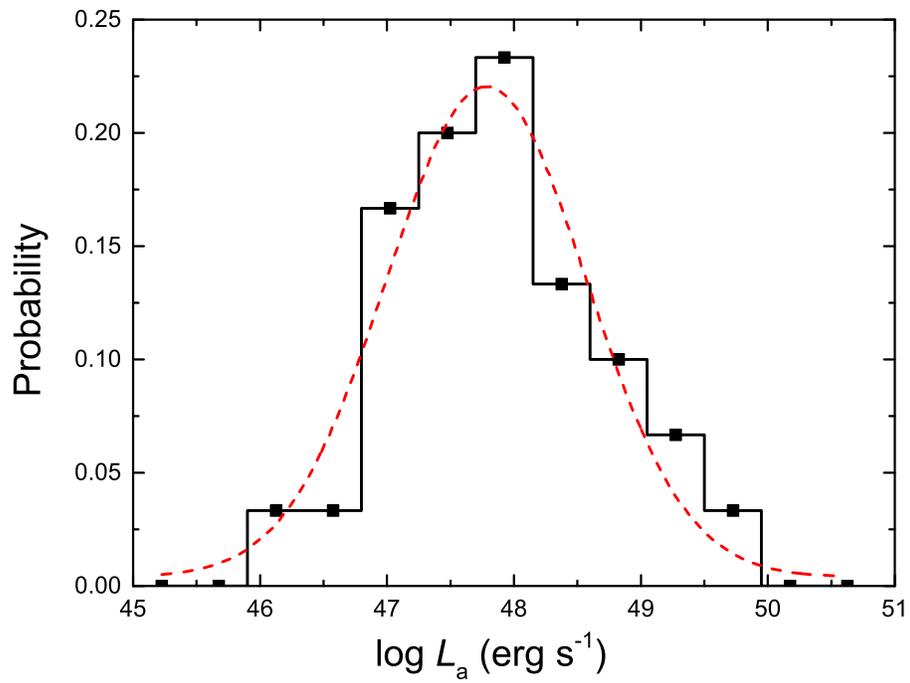}
\caption{Distribution of the X-ray afterglow luminosity at 3600 seconds post the BAT trigger for the GRBs with a single power-law decaying XRT lightcurve. The dashed line is the Gaussian fit to the distribution.}
\label{spl}
\end{figure}

It is difficult to discriminate the two possibilities with the current data\footnote{Yamazaki (2009) and Liang et al. (2009) proposed that the apparently difference may be due to the zero time effect to the canonical XRT lightcurves. They suggested the zero time of the canonical XRT lightcurve should be much prior the BAT trigger time.}. We make further simulation analysis for exploring whether the afterglow cover effect can lead to the observed $P_0-E_{\rm jet}$ relation. We search for GRBs that have a SPL XRT lightcurve from the current BAT GRB sample, and get 57 GRBs\footnote{The ratio of numbers of SPL lightcurves to canonical lightcurves in this analysis is 57/117=49\%, and it is 45\% for the redshift-known samples. This ratio is much larger than the reported in Liang et al. (2009). Note that this ratio highly depends on the selection for canonical lightcurves. We select only those canonical XRT lightcurves that have a good enough temporal coverage of the shallow-to-normal decay segment for our temporal and spectral analysis.}. Among them 30 GRBs have redshift measure. As shown in Figure \ref{Dis-L}, the typical $\tau$ value is about 3600 seconds. Therefore, we take the X-ray luminosity at 3600 seconds in the burst frame ($L_a$) as a reference to evaluate whether the MD wind emission of our mock GRBs is covered by the X-ray afterglows. We derive the $L_a$ distribution from the 30 GRBs, which is shown in Figure \ref{spl}. Weak X-ray afterglows with a low flux level may be covered by the MD emission. This may make a sharp cut-off at around $L_a=10^{47}$ erg s$^{-1}$ of the $L_a$ distribution. We fit the $L_{\rm a}$ distribution with a Gaussian function, which yields $\log L_a/{\rm erg\ s^{-1}}=47.78\pm 0.92$ (1$\sigma$). We also show the $L_a$ data of these GRBs in Figure \ref{threshold}(b). One can observe that the XRT detection has negligible effects in the detection of the plateau phase. The major factor hampering the identification of the plateau phase is the forward shock afterglow, not the XRT sensitivity. To evaluate whether the MD wind emission of a simulated GRB is covered by its X-ray afterglows, we generate an $L_a$ value from the Gaussian fit of the $L_a$ distribution via the bootstrap algorithm, and compare it with $L_b$ of a given GRB in the observed sample. If $L_{a}\geq L_{b}$, the MD wind emission is not detectable. The result is also shown in Figure \ref{simulation}. We find that the MD emission of $\sim 60\%$ mock GRBs is covered by their X-ray afterglows.

We do not find any correlation between $\log P_0$ and $\log E_{\rm jet}$ for the mock GRB sample. We measure the consistency of the final mock GRB sample with the observed one in the $\log P_0-\log E_{\rm jet}$ plane with the K-S test and get a $p$ value of $1.55\times 10^{-9}$. The K-S test indicates that the null hypothesis that the two samples are from the same parents can be rejected. Note that above simulation analysis is based on the assumption that the powers of the jet and MD wind injected by the magnetar are independent. The observational biases would not be the reasons that result in the observed $\log P_0-\log E_{\rm jet}$ relation. It may imply an intrinsic correlation between powers of the jet and MD wind. This is also reasonable since the powers of the jet and MD wind are extracted from the rotation of the magnetar. However, we should note that $E_{\rm jet}$ of the observed sample are clustered at $\log E_{\rm jet}/\rm erg=52\sim 53$ (Figure \ref{Dis-L}) and the derived correlation is driven by a few GRBs with $\log E_{\rm jet}/\rm erg<51$. Sub-energetic GRBs with detection of an X-ray plateau would be valuable for claiming it.

\subsection{Orphan MD emission and GW radiation of newly-born magnetars}
By comparing the observations of GRBs 101225A and 170714A to the MD wind emission of GRBs in our sample, we suggest that the prompt emission of the two GRBs may be dominated by long-lasting MD wind emission of magnetars, but not by ultra-long prompt emission of jets. The orphan MD emission would be due to the viewing angle effect. The misalignment of their jets to the light of sight may result in non-detection or weak detection of the jet prompt emission. For example, missing the jet emission of GRBs shown in Figure \ref{ultra-long}, their orphan X-ray plateau may mimic as an ultra-long GRBs analogue to GRBs 101225A and 170714A. They are different from the lightcurves of typical ultra-long GRBs. The prompt emission lightcurves of typical ultra-long GRBs are composed of substantial flares/pulses up to several thousand seconds even hours, such as that observed in GRBs 130925A (Piro et al. 2014), 121027A (Wu et al. 2013) and 111209A (Gendre et al. 2013). These ultra-long GRBs may be a population from collapses of supergiant progenitors but not a Wolf-Rayet progenitor as typical long GRBs (e.g., Woosley \& Heger 2012; Nakauchi et al. 2013; Gendre et al. 2013; Peng et al. 2013; Wu et al. 2013; Stratta et al. 2013; Levan et al. 2014; Virgili et al. 2013; Bo\"{e}r et al. 2015; Gao \& M\'{e}sz\'{a}ros 2015). The outer layers of such a progenitor may have sufficient angular momentum to form a disk for powering a long-lasting jet (e.g., Woosley \& Heger 2012).

Note that GRB 111209A was also suggested as an ultra-long GRB powered by magnetars (Greiner et al. 2015). It is occasionally discovered by {\em Swift}/BAT when it is settled to a pre-planned target. The mask-weighted BAT light curve shows an excess rate already around 150 seconds prior to the trigger time (Palmer et al. 2011). The start time of this event is missed by BAT. Observations of this GRB with Konus-Wind show a light curve with multi-peaked episode of emission from 5400 seconds prior to the BAT trigger time and $10^{4}$ seconds post the BAT trigger, making it as an exceptionally long GRB (Golenetskii et al. 2011). This event was active in its prompt phase for about 25000 seconds, making it the longest burst ever observed (Gendre et al. 2013). A supernova (2011kl) associated with GRB 111209A was observed (Greiner et al. 2015). The high luminosity and low metal-line opacity of the supernova suggest a scenario that extra energy is injected to power the supernova, favoring the idea that its central engine is a magnetar (Greiner et al. 2015). Interestingly, a long-lasting X-ray plateau is observed in its XRT lightcurve, similar to that of GRB 101225A (see Figure 1 in Gendre et al. 2013). The X-ray plateau may be dominated by the MD radiations of the magnetar. Although the progenitors of ultra-long GRBs may be different from the typical long GRBs, their central engines may be still similar. Because GRB jets are highly beamed, the detection rate of orphan MD wind emission would be higher than the jet emission with 2-3 orders of magnitude with an X-ray instrument being sensitive as XRT.
This may open a new approach for surveying X-ray selected magnetars with X-ray instruments, such as the Chinese-France Space Variable Object Monitors (SVOM; Wei et al. 2016) and the Einstein Probe (EP; Yuan et al. 2018).

\begin{figure}
\includegraphics[angle=0,scale=0.3]{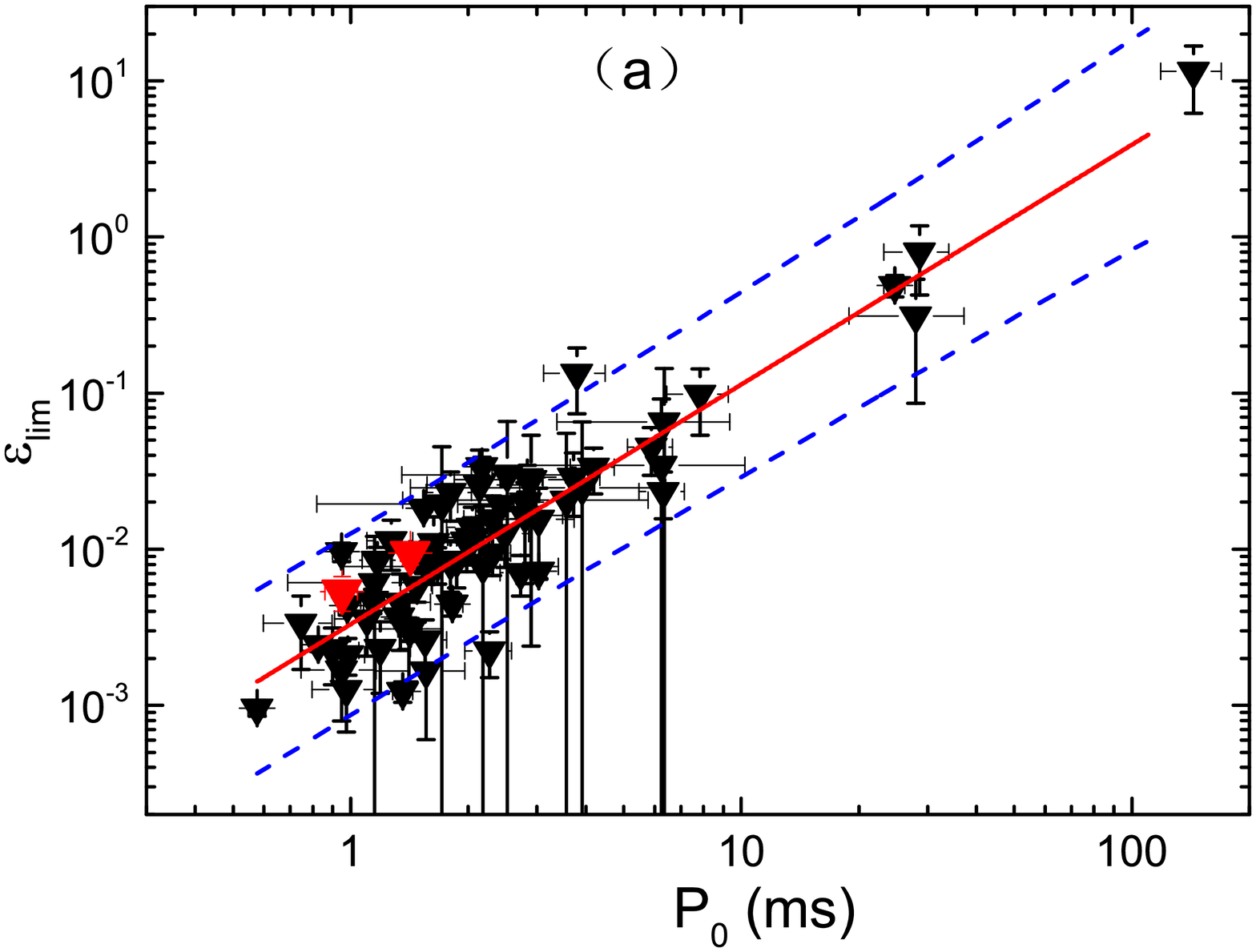}
\includegraphics[angle=0,scale=0.3]{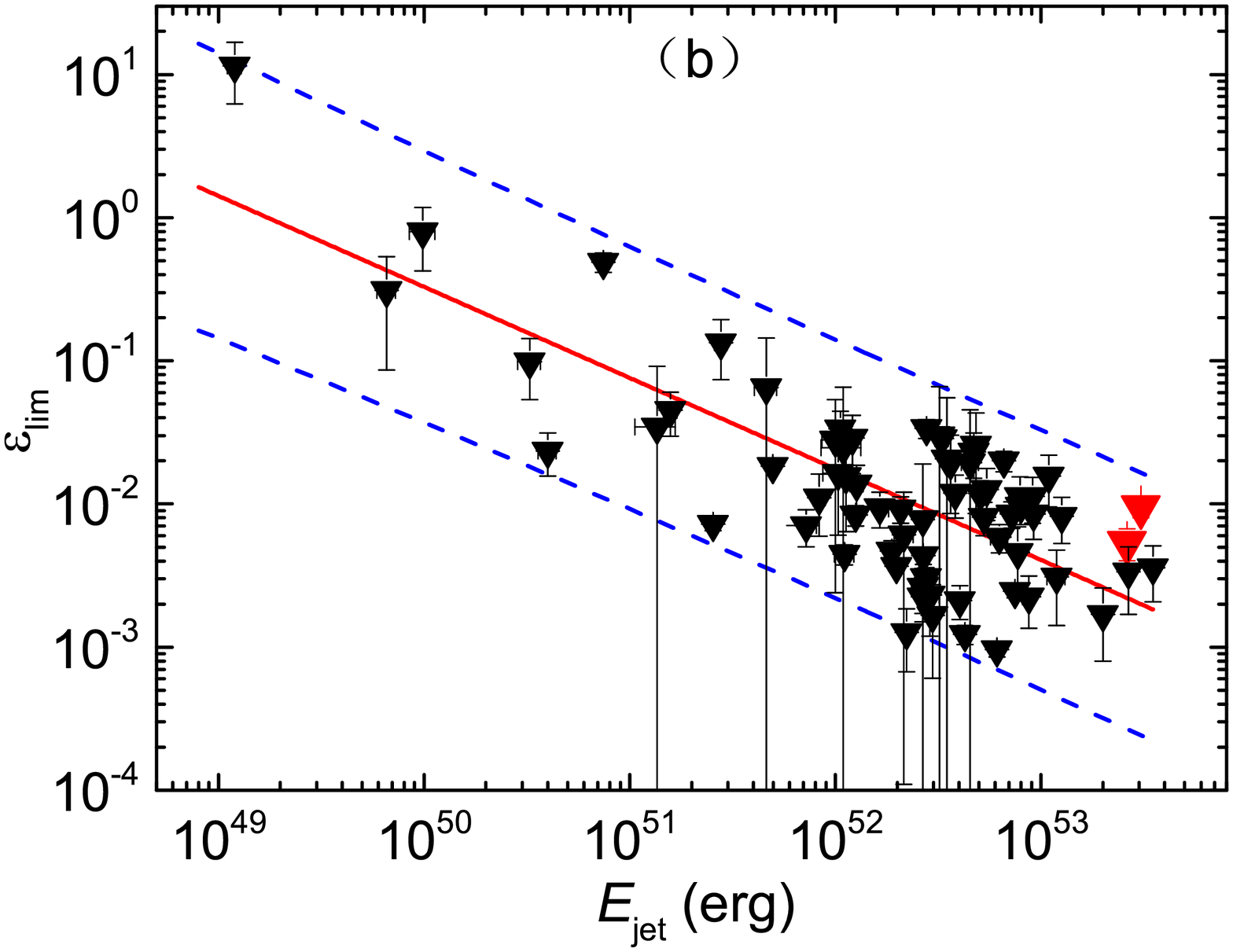}
\caption{$\varepsilon_{\rm lim}$ against $P_0$ and $E_{\rm jet}$ for the redshift-known GRBs in our sample. Solid and dashed lines are the least square linear fits and their 95\% confidence level, respectively. Red triangles are for GRBs 101225A and 170714A.}
\label{epsilon}
\end{figure}

Since the MD winds from long GRBs may be coasted by the supernova envelops, such surveys may be substantial for searching the MD wind emission of newly-born magnetars in compact star mergers (e.g., Gao et al. 2013). It was suggested that GRB 101225A may be from compact star merger (Th\"{o}ne et al. 2011) and the central engine of short GRBs 130603B may be also a supramassive magnetar (Fan et al. 2013). The GW emission from the mergers may be detectable with the advanced Laser Interferometer Gravitational-wave Observatory (aLIGO)/Virgo detectors, such as GW 170817 accompanying GRB 170817A (Abbott et al. 2017a, b; Savchenko et al. 2017). Since aLIGO can detect sources within $\sim 300$ Mpc only. In such a small observational volume the compact merger events are rare. Sensitive X-ray detectors for catching the wind emission may be helpful for increasing the detection possibility of electromagnetic counterparts of gravitational waves. In addition, GW observations for newly-born magnetars are constraining their properties since their GW luminosity is sensitive to its $P_0$ and ellipticity ($\varepsilon$), i.e., $L_{\rm GW}\propto \varepsilon^2 P_0^{-6}$ (e.g., Shapiro \& Teukolsky 1983; Zhang \& M\'{e}sz\'{a}ro 2001). A magnetar with larger $\varepsilon$ and faster rotation could radiate a stronger $L_{\rm GW}$. We calculate the upper limit of ellipticity for the magnetars in our sample with  (Fan et al. 2013; Lasky \& Glampedakis 2016; L\"{u} et al. 2017)
\begin{equation}
\varepsilon_{\rm lim} {=(\frac{15c^{5}\eta^2 I}{512G {L_{k}}^{2} \tau^{3}} \biggr)}^{1/2}=0.33\eta {I_{45}}^{1/2} {L_{k,49}}^{-1} {\tau_{2}}^{-3/2}
\end{equation}
where $G$ and $c$ are the gravitational constant and light speed, respectively. The results are presented in Table 2. We show $\varepsilon_{\rm lim}$ against $E_{\rm jet}$ and $\varepsilon_{\rm lim}-P_{0}$ in Figure \ref{epsilon}. One can observe a trend that a magnetar with a smaller $\varepsilon_{\rm lim}$ may be rotated faster and power a more energetic GRB. We made best linear fit with the maximum least quare method and get $\varepsilon_{\rm lim}\propto E^{-0.36\pm0.07}_{\rm jet,iso}$ and $\varepsilon_{\rm lim}\propto P_0^{1.52\pm0.11}$. If such an $\varepsilon-P_{0}$ relation is true, one can get $L_{\rm GW}\propto P_0^{\sim -3}$. Therefore, both GW and electromagnetic emission may highly depend on $P_0$.

\section{Summary}
Assuming that the early shallow decay segments of canonical XRT lightcurves of {\em Swift}/BAT GRBs are attributed to emission from the MD winds injected by newly-born magnetars, we have estimated the parameters of the magnetars and investigate possible relations among these parameters and their relation to the jet and MD wind radiations. We summary our results as following.
\begin{itemize}
 \item By making an extensive search from current {\em Swift/BAT} GRBs, we got a sample of 117 GRBs whose XRT lightcurves are canonical. Among them 67 GRBs have redshift measure. Their joint X-ray lightcurves derived from the BAT and XRT observations are well separated into the jet and wind emission epochs, and the shallow-to-normal decay segments have a good temporal coverage. We made temporal and spectral analysis for the BAT and XRT data and obtained the $E_{\rm jet}$, $E_{\rm wind}$, $L_b$, and $\tau$ values of these GRBs.
 \item The derived parameters of the magnetars of these GRBs are $P_0\in(0.6, 144.1)$ ms, $B_p\in(0.26, 22.40)\times 10^{15}$ G. A tightly correlation between $P_0$ and $E_{\rm jet}$ is found, i.e., $P_0\propto E^{-0.36\pm 0.03}_{\rm jet, iso}$. The $P_0-E_{\rm jet}$ relation reveals the connection between the jet prompt emission and properties of the GRB central engines. Since the GRB jets are collimated, we have $P_0\propto E^{-0.36}_{\rm jet}\theta_{\rm jet}^{0.76}$ based the $P_0-E_{\rm jet}$ relation. This hints that a magnetar with lower rotating speed may power a jet with smaller energetic and wider opening angle, if $E_{\rm jet}$ and $\theta_{\rm jet}$ are independent.

\item We have showed that the energy releases of the jets and winds are tightly correlated, i.e., $E_{\rm wind}\propto E^{0.91\pm 0.07}_{\rm jet, iso}$. This may indicate that the energy partition between the jet and wind among these GRBs are quasi-universal. Considering geometrical effect of the GRB jets, the energy partition between the jet and MD wind may be comparable.

\item  In the jet+wind paradigm for GRBs driven by magnetars, we have suggested that GRBs 101225A and GRB 170714A like GRBs, whose prompt gamma-ray/X-ray lightcurves are steady and long-lasting with a sharp drop,  are likely dominated by the orphan MD wind emission being due to misalignment to their bright jets. They may be distinct from flares/pulses-dominated ultra-long GRBs, which were proposed to be produced by different progenitors from that for typical long GRBs.
\end{itemize}
Our results are based on a sample of those GRBs that have a shallow decay segment in their early XRT lightcurves. This sample suffers the observational biases of BAT and XRT fluctuation thresholds. In addition, the shallow-decay segment may be also covered by bright jet afterglow emission, leading to detection of a SPL afterglow lightcurve only. We present simulation analysis for evaluating whether our analysis results are resulted from these observational biases. We show that the these observational biases only cannot make the $\log P_0-\log E_{\rm jet}$ relation. However, we should emphasize this relation is driven by a few sub energetic GRBs. Sub-energetic GRBs with detection of an X-ray plateau would be valuable for confirmation of our results. Discussion on orphan MD emission and GW radiation of newly-born magnetars is also presented.
\acknowledgments
We thanks the anonymous referee for his/her valuable comments and suggestions. We acknowledge the use of the public data from the {\em Swift} data archive and the UK {\em Swift} Science Data Center. This work is supported by the National Natural Science Foundation of China (Grant No.11533003, 11851304, 11603006, and U1731239), Guangxi Science Foundation (grant No. 2017GXNSFFA198008, 2016GXNSFCB380005 and AD17129006), the One-Hundred-Talents Program of Guangxi colleges, the high level innovation team and outstanding scholar program in Guangxi colleges, Scientific Research Foundation of Guangxi University
(grant No. XGZ150299), and special funding for Guangxi distinguished professors (2017AD22006).

\begin{center}
\begin{deluxetable}{lllllllllllll}
\rotate \tablewidth{0pt} \tabletypesize{\footnotesize}
\tablecaption{Observational Properties of the GRBs in our sample}\tablenum{1} \tablehead{
\colhead{GRB}& \colhead{$z$}& \colhead{$\alpha_{1}$\tablenotemark{a}}&
\colhead{$\alpha_{2}$\tablenotemark{a}}& \colhead{$\log t_{\rm b}$\tablenotemark{a}}& \colhead{$\log
T_1-\log T_2$\tablenotemark{b}}& \colhead{$\beta_{\rm jet}$\tablenotemark{c}}& \colhead{$\beta_{\rm
wind}$\tablenotemark{c}}& \colhead{$\log L_{\rm b,45}$\tablenotemark{d} }& \colhead{$\log E_{\rm
jet, 50}$\tablenotemark{e} }& \colhead{$\log E_{\rm wind,50}$ \tablenotemark{e}}& }

\startdata
\hline
050315	&	1.949	&	0.26 	$\pm$	0.04 	&	1.34 	$\pm$	0.09 	&	4.81 	$\pm$	3.99 	&	3.72 	-	5.77 	&	1.11 	$\pm$	0.09 	&	0.85 	$\pm$	0.07 	&	2.06 	$\pm$	1.13 	&	2.43 	$\pm$	1.06 	&	1.29 	&	\\
050319	&	3.24	&	0.59 	$\pm$	0.06 	&	1.67 	$\pm$	0.13 	&	4.62 	$\pm$	3.89 	&	2.78 	-	5.73 	&	1.07 	$\pm$	0.20 	&	0.89 	$\pm$	0.08 	&	2.64 	$\pm$	1.95 	&	2.46 	$\pm$	1.50 	&	1.35 	&	\\
050713A	&		&	0.49 	$\pm$	0.13 	&	1.26 	$\pm$	0.04 	&	3.92 	$\pm$	3.38 	&	3.04 	-	6.22 	&	0.55 	$\pm$	0.08 	&	1.08 	$\pm$	0.15 	&				&				&		&	\\
050713B	&		&	-0.17 	$\pm$	0.08 	&	1.05 	$\pm$	0.03 	&	4.09 	$\pm$	3.12 	&	3.01 	-	5.73 	&	0.39 	$\pm$	0.17 	&	1.04 	$\pm$	0.19 	&				&				&		&	\\
050802	&	1.71	&	0.63 	$\pm$	0.04 	&	1.69 	$\pm$	0.05 	&	3.81 	$\pm$	2.91 	&	2.53 	-	5.03 	&	0.52 	$\pm$	0.15 	&	0.65 	$\pm$	0.11 	&	2.78 	$\pm$	1.97 	&	2.10 	$\pm$	0.98 	&	0.89 	&	\\
050814	&	5.3	&	0.56 	$\pm$	0.04 	&	1.81 	$\pm$	0.24 	&	4.84 	$\pm$	4.17 	&	3.05 	-	5.68 	&	0.83 	$\pm$	0.18 	&	0.89 	$\pm$	0.11 	&	2.45 	$\pm$	1.78 	&	3.08 	$\pm$	2.08 	&	1.44 	&	\\
050822	&	1.434	&	0.23 	$\pm$	0.07 	&	1.04 	$\pm$	0.03 	&	4.16 	$\pm$	3.45 	&	3.01 	-	6.63 	&	1.44 	$\pm$	0.15 	&	0.76 	$\pm$	0.18 	&	1.87 	$\pm$	1.04 	&	2.08 	$\pm$	0.87 	&	-0.10 	&	\\
050826	&	0.297	&	0.13 	$\pm$	0.20 	&	1.82 	$\pm$	0.19 	&	4.54 	$\pm$	3.81 	&	4.06 	-	5.14 	&	0.14 	$\pm$	0.30 	&	0.91 	$\pm$	0.49 	&	-0.55 	$\pm$	-1.30 	&	0.00 	$\pm$	-0.86 	&	-2.00 	&	\\
050915B	&		&	0.45 	$\pm$	0.05 	&	1.60 	$\pm$	0.29 	&	5.01 	$\pm$	4.55 	&	3.12 	-	5.68 	&	0.90 	$\pm$	0.06 	&	1.19 	$\pm$	0.34 	&				&				&		&	\\
051016B	&	0.9364	&	0.38 	$\pm$	0.05 	&	1.27 	$\pm$	0.06 	&	4.30 	$\pm$	3.59 	&	2.72 	-	6.09 	&	1.40 	$\pm$	0.24 	&	0.79 	$\pm$	0.16 	&	0.99 	$\pm$	0.20 	&	0.51 	$\pm$	-0.39 	&	-0.27 	&	\\
060109	&		&	-0.10 	$\pm$	0.10 	&	1.49 	$\pm$	0.06 	&	3.78 	$\pm$	2.86 	&	2.92 	-	6.58 	&	0.93 	$\pm$	0.25 	&	1.13 	$\pm$	0.16 	&				&				&		&	\\
060202	&	0.785	&	0.29 	$\pm$	0.05 	&	6.88 	$\pm$	0.27 	&	2.89 	$\pm$	0.90 	&	2.49 	-	3.09 	&	0.61 	$\pm$	0.07 	&	1.03 	$\pm$	0.09 	&	3.77 	$\pm$	2.21 	&	1.70 	$\pm$	0.33 	&	1.27 	&	\\
060204B	&	2.3393	&	0.68 	$\pm$	0.06 	&	1.58 	$\pm$	0.06 	&	3.91 	$\pm$	3.24 	&	2.63 	-	5.53 	&	0.46 	$\pm$	0.09 	&	1.09 	$\pm$	0.12 	&	2.63 	$\pm$	2.01 	&	2.56 	$\pm$	1.30 	&	0.55 	&	\\
060211A	&		&	0.48 	$\pm$	0.07 	&	2.10 	$\pm$	1.45 	&	5.50 	$\pm$	5.12 	&	3.74 	-	5.76 	&	0.76 	$\pm$	0.12 	&	1.15 	$\pm$	0.32 	&				&				&		&	\\
060219	&		&	0.45 	$\pm$	0.15 	&	1.52 	$\pm$	0.14 	&	4.51 	$\pm$	4.07 	&	3.67 	-	5.74 	&	1.56 	$\pm$	0.36 	&	1.93 	$\pm$	0.38 	&				&				&		&	\\
060428A	&		&	0.53 	$\pm$	0.02 	&	1.44 	$\pm$	0.05 	&	4.99 	$\pm$	4.12 	&	2.36 	-	6.53 	&	1.04 	$\pm$	0.11 	&	0.89 	$\pm$	0.13 	&				&				&		&	\\
060502A	&	1.51	&	0.50 	$\pm$	0.04 	&	1.15 	$\pm$	0.04 	&	4.36 	$\pm$	3.76 	&	2.53 	-	6.14 	&	0.47 	$\pm$	0.08 	&	0.95 	$\pm$	0.14 	&	1.86 	$\pm$	1.18 	&	2.05 	$\pm$	0.69 	&	0.10 	&	\\
060510A	&		&	0.11 	$\pm$	0.05 	&	1.50 	$\pm$	0.03 	&	3.78 	$\pm$	2.74 	&	2.33 	-	5.71 	&	0.59 	$\pm$	0.07 	&	0.70 	$\pm$	0.11 	&				&				&		&	\\
060604	&	2.68	&	0.39 	$\pm$	0.10 	&	1.25 	$\pm$	0.06 	&	4.33 	$\pm$	3.76 	&	3.09 	-	5.85 	&	1.12 	$\pm$	0.45 	&	1.06 	$\pm$	0.12 	&	2.13 	$\pm$	1.49 	&	2.01 	$\pm$	1.19 	&	0.63 	&	\\
060605	&	3.8	&	0.48 	$\pm$	0.04 	&	2.08 	$\pm$	0.08 	&	3.95 	$\pm$	2.84 	&	2.55 	-	4.85 	&	0.47 	$\pm$	0.22 	&	1.17 	$\pm$	0.16 	&	3.12 	$\pm$	2.10 	&	2.32 	$\pm$	1.42 	&	1.07 	&	\\
060607A	&	3.082	&	0.37 	$\pm$	0.05 	&	3.67 	$\pm$	0.11 	&	4.10 	$\pm$	2.59 	&	2.64 	-	4.95 	&	0.47 	$\pm$	0.08 	&	0.52 	$\pm$	0.08 	&	3.77 	$\pm$	2.61 	&	2.79 	$\pm$	1.34 	&	1.98 	&	\\
060614	&	0.125	&	0.05 	$\pm$	0.05 	&	1.82 	$\pm$	0.04 	&	4.66 	$\pm$	3.42 	&	3.69 	-	6.25 	&	1.13 	$\pm$	0.04 	&	0.75 	$\pm$	0.10 	&	-0.60 	$\pm$	-1.70 	&	0.87 	$\pm$	-0.91 	&	-1.52 	&	\\
060712	&		&	0.41 	$\pm$	0.09 	&	1.17 	$\pm$	0.06 	&	4.03 	$\pm$	3.61 	&	2.70 	-	5.79 	&	0.66 	$\pm$	0.33 	&	1.66 	$\pm$	0.54 	&				&				&		&	\\
060714	&	2.71	&	0.12 	$\pm$	0.21 	&	1.24 	$\pm$	0.04 	&	3.48 	$\pm$	2.89 	&	2.72 	-	5.92 	&	0.98 	$\pm$	0.11 	&	0.80 	$\pm$	0.16 	&	3.22 	$\pm$	2.66 	&	2.69 	$\pm$	1.40 	&	0.71 	&	\\
060729	&	0.54	&	0.13 	$\pm$	0.02 	&	1.38 	$\pm$	0.01 	&	4.81 	$\pm$	3.40 	&	2.63 	-	6.79 	&	0.88 	$\pm$	0.14 	&	1.03 	$\pm$	0.11 	&	1.20 	$\pm$	-0.22 	&	1.41 	$\pm$	0.12 	&	0.40 	&	\\
060807	&		&	0.05 	$\pm$	0.04 	&	1.76 	$\pm$	0.05 	&	3.91 	$\pm$	2.75 	&	2.28 	-	5.57 	&	0.58 	$\pm$	0.21 	&	0.89 	$\pm$	0.12 	&				&				&		&	\\
060814	&	1.9229	&	0.41 	$\pm$	0.07 	&	1.38 	$\pm$	0.03 	&	4.07 	$\pm$	3.22 	&	3.09 	-	6.08 	&	0.56 	$\pm$	0.03 	&	0.94 	$\pm$	0.11 	&	2.67 	$\pm$	1.83 	&	3.10 	$\pm$	1.26 	&	0.95 	&	\\
061121	&	1.314	&	0.48 	$\pm$	0.03 	&	1.43 	$\pm$	0.02 	&	3.85 	$\pm$	2.86 	&	2.44 	-	6.28 	&	0.38 	$\pm$	0.03 	&	0.89 	$\pm$	0.16 	&	2.85 	$\pm$	1.90 	&	2.72 	$\pm$	0.86 	&	1.20 	&	\\
061202	&	2.253	&	-0.05 	$\pm$	0.09 	&	1.52 	$\pm$	0.07 	&	4.25 	$\pm$	3.32 	&	3.63 	-	5.62 	&	0.63 	$\pm$	0.07 	&	0.95 	$\pm$	0.11 	&	3.07 	$\pm$	2.02 	&	2.61 	$\pm$	1.15 	&	1.27 	&	\\
061222A	&	2.088	&	0.35 	$\pm$	0.02 	&	1.39 	$\pm$	0.01 	&	3.79 	$\pm$	2.51 	&	2.44 	-	6.14 	&	0.38 	$\pm$	0.04 	&	0.84 	$\pm$	0.06 	&	3.65 	$\pm$	1.92 	&	2.87 	$\pm$	1.15 	&	1.56 	&	\\
070110	&	2.352	&	0.01 	$\pm$	0.06 	&	7.88 	$\pm$	0.60 	&	4.30 	$\pm$	2.53 	&	3.64 	-	4.45 	&	0.83 	$\pm$	0.13 	&	1.03 	$\pm$	0.03 	&	2.77 	$\pm$	1.46 	&	2.30 	$\pm$	1.09 	&	1.09 	&	\\
070129	&	2.3384	&	0.24 	$\pm$	0.07 	&	1.16 	$\pm$	0.04 	&	4.31 	$\pm$	3.55 	&	3.14 	-	6.10 	&	1.04 	$\pm$	0.16 	&	1.16 	$\pm$	0.15 	&	2.21 	$\pm$	1.34 	&	2.74 	$\pm$	1.48 	&	1.30 	&	\\
070306	&	1.496	&	0.06 	$\pm$	0.05 	&	1.80 	$\pm$	0.05 	&	4.45 	$\pm$	3.28 	&	3.66 	-	5.96 	&	0.72 	$\pm$	0.10 	&	0.79 	$\pm$	0.10 	&	2.43 	$\pm$	1.25 	&	2.44 	$\pm$	1.13 	&	0.88 	&	\\
070328	&	2.0627	&	0.23 	$\pm$	0.05 	&	1.49 	$\pm$	0.15 	&	2.84 	$\pm$	1.59 	&	2.32 	-	5.88 	&	0.26 	$\pm$	0.00 	&	1.06 	$\pm$	0.04 	&	4.48 	$\pm$	3.21 	&	2.90 	$\pm$	1.20 	&	1.79 	&	\\
070420	&		&	0.34 	$\pm$	0.06 	&	1.49 	$\pm$	0.03 	&	3.61 	$\pm$	2.70 	&	2.49 	-	5.68 	&	0.59 	$\pm$	0.05 	&	0.99 	$\pm$	0.14 	&				&				&		&	\\
070508	&	0.82	&	0.48 	$\pm$	0.02 	&	1.46 	$\pm$	0.01 	&	2.97 	$\pm$	1.76 	&	1.97 	-	5.86 	&	0.36 	$\pm$	0.03 	&	0.59 	$\pm$	0.19 	&	3.41 	$\pm$	2.17 	&	2.44 	$\pm$	0.59 	&	1.23 	&	\\
070521	&	2.0865	&	0.15 	$\pm$	0.06 	&	1.57 	$\pm$	0.04 	&	3.31 	$\pm$	2.16 	&	2.07 	-	5.40 	&	0.38 	$\pm$	0.04 	&	0.74 	$\pm$	0.29 	&	3.83 	$\pm$	2.84 	&	2.86 	$\pm$	1.20 	&	1.18 	&	\\
070616	&		&	-0.14 	$\pm$	0.04 	&	4.73 	$\pm$	0.11 	&	2.70 	$\pm$	0.70 	&	2.15 	-	3.04 	&	0.59 	$\pm$	0.12 	&	1.05 	$\pm$	0.08 	&				&				&		&	\\
080229A	&		&	0.15 	$\pm$	0.06 	&	1.26 	$\pm$	0.02 	&	3.37 	$\pm$	2.34 	&	2.38 	-	5.71 	&	0.90 	$\pm$	0.06 	&	0.74 	$\pm$	0.12 	&				&				&		&	\\
080310	&	2.43	&	0.19 	$\pm$	0.07 	&	1.66 	$\pm$	0.06 	&	4.06 	$\pm$	3.00 	&	3.12 	-	5.66 	&	1.32 	$\pm$	0.16 	&	0.92 	$\pm$	0.14 	&	2.63 	$\pm$	1.62 	&	2.70 	$\pm$	1.39 	&	0.60 	&	\\
080430	&	0.767	&	0.40 	$\pm$	0.02 	&	1.12 	$\pm$	0.03 	&	4.44 	$\pm$	3.61 	&	2.70 	-	6.36 	&	0.74 	$\pm$	0.09 	&	0.98 	$\pm$	0.11 	&	1.06 	$\pm$	0.12 	&	1.20 	$\pm$	0.09 	&	-0.05 	&	\\
080905B	&	2.374	&	0.33 	$\pm$	0.14 	&	1.44 	$\pm$	0.05 	&	3.62 	$\pm$	3.19 	&	2.39 	-	5.89 	&	0.76 	$\pm$	0.14 	&	0.70 	$\pm$	0.13 	&	3.57 	$\pm$	3.20 	&	2.33 	$\pm$	1.37 	&	1.06 	&	\\
081029	&	3.847	&	0.44 	$\pm$	0.07 	&	2.82 	$\pm$	0.18 	&	4.25 	$\pm$	3.12 	&	3.45 	-	5.46 	&	0.44 	$\pm$	0.18 	&	0.85 	$\pm$	0.14 	&	2.88 	$\pm$	1.93 	&	2.80 	$\pm$	1.78 	&	1.22 	&	\\
081126	&		&	0.27 	$\pm$	0.19 	&	1.50 	$\pm$	0.06 	&	3.52 	$\pm$	3.19 	&	2.23 	-	5.61 	&	0.21 	$\pm$	0.06 	&	0.56 	$\pm$	0.27 	&				&				&		&	\\
081128	&		&	0.29 	$\pm$	0.36 	&	1.41 	$\pm$	0.19 	&	4.35 	$\pm$	4.09 	&	3.65 	-	5.68 	&	0.98 	$\pm$	0.09 	&	0.33 	$\pm$	0.61 	&				&				&		&	\\
090404	&	3	&	0.23 	$\pm$	0.05 	&	1.02 	$\pm$	0.05 	&	4.20 	$\pm$	3.57 	&	2.56 	-	6.10 	&	1.32 	$\pm$	0.09 	&	1.21 	$\pm$	0.15 	&	2.94 	$\pm$	2.14 	&	2.89 	$\pm$	1.27 	&	1.19 	&	\\
090407	&	1.4485	&	0.36 	$\pm$	0.03 	&	1.63 	$\pm$	0.09 	&	4.84 	$\pm$	3.94 	&	3.09 	-	5.95 	&	0.76 	$\pm$	0.30 	&	1.24 	$\pm$	0.13 	&	1.46 	$\pm$	0.48 	&	1.86 	$\pm$	0.94 	&	0.71 	&	\\
090516A	&	4.109	&	0.62 	$\pm$	0.11 	&	1.84 	$\pm$	0.07 	&	4.18 	$\pm$	3.36 	&	3.57 	-	5.42 	&	0.84 	$\pm$	0.11 	&	0.95 	$\pm$	0.10 	&	3.24 	$\pm$	2.53 	&	3.55 	$\pm$	2.31 	&	1.40 	&	\\
090529	&	2.625	&	0.19 	$\pm$	0.30 	&	1.04 	$\pm$	0.14 	&	4.55 	$\pm$	4.43 	&	3.70 	-	5.91 	&	1.06 	$\pm$	0.18 	&	0.52 	$\pm$	0.24 	&	1.61 	$\pm$	1.33 	&	2.04 	$\pm$	1.38 	&	-0.25 	&	\\
090618	&	0.54	&	0.65 	$\pm$	0.02 	&	1.46 	$\pm$	0.01 	&	3.80 	$\pm$	2.60 	&	2.86 	-	6.50 	&	0.71 	$\pm$	0.02 	&	0.95 	$\pm$	0.03 	&	2.28 	$\pm$	1.13 	&	2.82 	$\pm$	0.79 	&	0.87 	&	\\
090727	&		&	0.55 	$\pm$	0.05 	&	1.72 	$\pm$	0.33 	&	5.70 	$\pm$	5.24 	&	3.63 	-	6.32 	&	0.24 	$\pm$	0.24 	&	0.91 	$\pm$	0.28 	&				&				&		&	\\
090728	&		&	0.25 	$\pm$	0.13 	&	1.87 	$\pm$	0.12 	&	3.38 	$\pm$	2.60 	&	2.41 	-	4.91 	&	1.05 	$\pm$	0.27 	&	0.76 	$\pm$	0.15 	&				&				&		&	\\
090813	&		&	0.19 	$\pm$	0.04 	&	1.27 	$\pm$	0.01 	&	2.74 	$\pm$	1.62 	&	1.96 	-	5.81 	&	0.69 	$\pm$	0.12 	&	1.02 	$\pm$	0.08 	&				&				&		&	\\
090904A	&		&	0.22 	$\pm$	0.11 	&	1.40 	$\pm$	0.10 	&	4.20 	$\pm$	3.54 	&	3.03 	-	5.50 	&	1.01 	$\pm$	0.10 	&	1.26 	$\pm$	0.43 	&				&				&		&	\\
091018	&	0.971	&	0.36 	$\pm$	0.07 	&	1.24 	$\pm$	0.02 	&	2.78 	$\pm$	2.05 	&	1.82 	-	5.86 	&	1.30 	$\pm$	0.06 	&	0.73 	$\pm$	0.13 	&	3.15 	$\pm$	2.37 	&	1.44 	$\pm$	0.30 	&	0.48 	&	\\
091029	&	2.752	&	0.22 	$\pm$	0.05 	&	1.15 	$\pm$	0.03 	&	4.08 	$\pm$	3.24 	&	2.92 	-	6.28 	&	0.88 	$\pm$	0.07 	&	1.11 	$\pm$	0.12 	&	2.66 	$\pm$	1.72 	&	2.58 	$\pm$	1.19 	&	0.82 	&	\\
091130B	&		&	0.30 	$\pm$	0.06 	&	1.21 	$\pm$	0.07 	&	4.76 	$\pm$	4.09 	&	3.60 	-	6.06 	&	1.10 	$\pm$	0.15 	&	1.27 	$\pm$	0.16 	&				&				&		&	\\
100302A	&	4.813	&	0.47 	$\pm$	0.08 	&	1.08 	$\pm$	0.12 	&	4.76 	$\pm$	4.56 	&	3.18 	-	6.00 	&	0.81 	$\pm$	0.20 	&	0.79 	$\pm$	0.22 	&	2.12 	$\pm$	1.80 	&	2.43 	$\pm$	1.26 	&	0.79 	&	\\
100305A	&		&	0.56 	$\pm$	0.20 	&	2.04 	$\pm$	0.16 	&	4.10 	$\pm$	3.45 	&	3.56 	-	5.33 	&	0.27 	$\pm$	0.23 	&	0.97 	$\pm$	0.18 	&				&				&		&	\\
100418A	&	0.6235	&	-0.09 	$\pm$	0.06 	&	1.67 	$\pm$	0.10 	&	5.08 	$\pm$	4.21 	&	3.00 	-	6.32 	&	1.16 	$\pm$	0.25 	&	0.86 	$\pm$	0.29 	&	0.31 	$\pm$	-0.57 	&	0.60 	$\pm$	-0.38 	&	-1.10 	&	\\
100425A	&	1.755	&	0.48 	$\pm$	0.06 	&	1.19 	$\pm$	0.15 	&	4.44 	$\pm$	4.17 	&	2.68 	-	5.67 	&	1.41 	$\pm$	0.29 	&	1.18 	$\pm$	0.25 	&	1.18 	$\pm$	0.79 	&	1.66 	$\pm$	0.76 	&	0.00 	&	\\
100508A	&		&	0.35 	$\pm$	0.03 	&	2.98 	$\pm$	0.24 	&	4.43 	$\pm$	3.30 	&	2.79 	-	5.32 	&	0.19 	$\pm$	0.24 	&	0.35 	$\pm$	0.14 	&				&				&		&	\\
100522A	&		&	0.46 	$\pm$	0.05 	&	1.34 	$\pm$	0.08 	&	4.34 	$\pm$	3.67 	&	2.87 	-	5.58 	&	0.92 	$\pm$	0.09 	&	1.23 	$\pm$	0.14 	&				&				&		&	\\
100614A	&		&	0.43 	$\pm$	0.05 	&	2.35 	$\pm$	0.31 	&	5.23 	$\pm$	4.36 	&	3.74 	-	5.75 	&	0.88 	$\pm$	0.15 	&	1.08 	$\pm$	0.22 	&				&				&		&	\\
100615A	&	1.398	&	0.36 	$\pm$	0.03 	&	1.45 	$\pm$	0.17 	&	4.55 	$\pm$	3.87 	&	2.32 	-	5.22 	&	0.87 	$\pm$	0.04 	&	1.21 	$\pm$	0.17 	&	2.64 	$\pm$	1.82 	&	2.35 	$\pm$	0.61 	&	1.01 	&	\\
100704A	&	3.6	&	0.70 	$\pm$	0.03 	&	1.36 	$\pm$	0.05 	&	4.53 	$\pm$	3.86 	&	2.73 	-	6.16 	&	0.73 	$\pm$	0.06 	&	1.04 	$\pm$	0.09 	&	2.97 	$\pm$	2.29 	&	3.30 	$\pm$	1.72 	&	1.71 	&	\\
100725B	&		&	0.36 	$\pm$	0.07 	&	1.39 	$\pm$	0.07 	&	4.24 	$\pm$	3.48 	&	2.91 	-	5.41 	&	0.89 	$\pm$	0.06 	&	1.46 	$\pm$	0.23 	&				&				&		&	\\
100814A	&	1.44	&	0.47 	$\pm$	0.02 	&	2.07 	$\pm$	0.07 	&	5.16 	$\pm$	3.94 	&	3.61 	-	6.22 	&	0.47 	$\pm$	0.04 	&	0.82 	$\pm$	0.05 	&	1.74 	$\pm$	0.52 	&	2.63 	$\pm$	0.94 	&	1.08 	&	\\
100901A	&	1.408	&	-0.10 	$\pm$	0.06 	&	1.48 	$\pm$	0.05 	&	4.54 	$\pm$	3.36 	&	3.67 	-	6.20 	&	0.52 	$\pm$	0.20 	&	1.04 	$\pm$	0.42 	&	2.11 	$\pm$	0.89 	&	2.05 	$\pm$	1.09 	&	0.80 	&	\\
100906A	&	1.727	&	0.73 	$\pm$	0.05 	&	2.09 	$\pm$	0.08 	&	4.08 	$\pm$	3.15 	&	2.65 	-	5.30 	&	0.84 	$\pm$	0.04 	&	0.89 	$\pm$	0.11 	&	2.60 	$\pm$	1.81 	&	2.96 	$\pm$	0.00 	&	1.14 	&	\\
101024A	&		&	0.03 	$\pm$	0.08 	&	1.36 	$\pm$	0.03 	&	3.03 	$\pm$	2.02 	&	2.06 	-	5.15 	&	0.84 	$\pm$	0.07 	&	1.02 	$\pm$	0.23 	&				&				&		&	\\
110102A	&		&	0.48 	$\pm$	0.03 	&	1.46 	$\pm$	0.03 	&	4.13 	$\pm$	3.05 	&	2.75 	-	5.92 	&	0.60 	$\pm$	0.04 	&	1.04 	$\pm$	0.09 	&				&				&		&	\\
110213A	&	1.46	&	0.00 	$\pm$	0.05 	&	1.81 	$\pm$	0.02 	&	3.49 	$\pm$	2.16 	&	2.36 	-	5.68 	&	0.83 	$\pm$	0.12 	&	0.77 	$\pm$	0.12 	&	3.70 	$\pm$	2.52 	&	2.43 	$\pm$	1.25 	&	1.61 	&	\\
110420A	&		&	0.31 	$\pm$	0.06 	&	1.24 	$\pm$	0.03 	&	3.66 	$\pm$	2.88 	&	2.34 	-	6.08 	&	1.33 	$\pm$	0.07 	&	0.46 	$\pm$	0.20 	&				&				&		&	\\
110808A	&	1.348	&	0.42 	$\pm$	0.11 	&	1.12 	$\pm$	0.20 	&	4.89 	$\pm$	4.76 	&	3.63 	-	5.84 	&	1.32 	$\pm$	0.40 	&	1.32 	$\pm$	0.37 	&	0.68 	$\pm$	0.41 	&	1.13 	$\pm$	0.48 	&	-0.52 	&	\\
110820A	&		&	0.31 	$\pm$	0.10 	&	1.66 	$\pm$	0.36 	&	4.74 	$\pm$	4.26 	&	3.00 	-	5.46 	&	0.98 	$\pm$	0.27 	&	1.36 	$\pm$	0.42 	&				&				&		&	\\
111008A	&	4.9898	&	0.09 	$\pm$	0.10 	&	1.22 	$\pm$	0.03 	&	3.64 	$\pm$	2.93 	&	2.61 	-	5.96 	&	0.86 	$\pm$	0.09 	&	0.94 	$\pm$	0.15 	&	4.18 	$\pm$	3.48 	&	3.42 	$\pm$	2.17 	&	1.49 	&	\\
111228A	&	0.714	&	0.37 	$\pm$	0.04 	&	1.26 	$\pm$	0.03 	&	4.09 	$\pm$	3.23 	&	2.84 	-	6.22 	&	1.27 	$\pm$	0.06 	&	0.94 	$\pm$	0.13 	&	1.69 	$\pm$	0.77 	&	2.03 	$\pm$	0.33 	&	0.19 	&	\\
120118B	&	2.943	&	0.19 	$\pm$	0.20 	&	1.08 	$\pm$	0.11 	&	3.66 	$\pm$	3.39 	&	2.87 	-	4.90 	&	1.04 	$\pm$	0.12 	&	1.02 	$\pm$	0.27 	&	2.91 	$\pm$	2.52 	&	2.51 	$\pm$	1.25 	&	0.33 	&	\\
120308A	&		&	0.65 	$\pm$	0.03 	&	2.41 	$\pm$	0.12 	&	4.11 	$\pm$	3.04 	&	2.59 	-	5.23 	&	0.71 	$\pm$	0.13 	&	0.46 	$\pm$	0.10 	&				&				&		&	\\
120324A	&		&	0.23 	$\pm$	0.09 	&	1.03 	$\pm$	0.04 	&	3.60 	$\pm$	3.07 	&	2.54 	-	5.13 	&	0.34 	$\pm$	0.04 	&	1.12 	$\pm$	0.27 	&				&				&		&	\\
120422A	&	0.283	&	0.29 	$\pm$	0.04 	&	1.26 	$\pm$	0.36 	&	5.47 	$\pm$	5.08 	&	2.69 	-	5.97 	&	0.27 	$\pm$	0.24 	&	1.06 	$\pm$	0.39 	&	-1.52 	$\pm$	-2.00 	&	-0.18 	$\pm$	-1.17 	&	-2.00 	&	\\
120521C	&	6	&	0.36 	$\pm$	0.08 	&	2.53 	$\pm$	0.45 	&	4.34 	$\pm$	3.50 	&	3.25 	-	4.58 	&	0.73 	$\pm$	0.11 	&	0.78 	$\pm$	0.32 	&	2.70 	$\pm$	1.88 	&	2.90 	$\pm$	1.84 	&	0.43 	&	\\
120811C	&	2.671	&	0.40 	$\pm$	0.28 	&	1.21 	$\pm$	0.10 	&	3.33 	$\pm$	3.08 	&	2.48 	-	4.92 	&	1.04 	$\pm$	0.06 	&	0.65 	$\pm$	0.14 	&	3.56 	$\pm$	3.24 	&	2.66 	$\pm$	1.64 	&	0.19 	&	\\
121027A	&	1.773	&	0.39 	$\pm$	0.17 	&	1.52 	$\pm$	0.08 	&	5.19 	$\pm$	4.58 	&	4.53 	-	6.42 	&	0.82 	$\pm$	0.09 	&	1.22 	$\pm$	0.17 	&	1.65 	$\pm$	1.09 	&	2.47 	$\pm$	0.82 	&	0.15 	&	\\
121217A	&		&	0.32 	$\pm$	0.06 	&	1.35 	$\pm$	0.04 	&	4.29 	$\pm$	3.41 	&	3.21 	-	6.06 	&	0.53 	$\pm$	0.08 	&	1.01 	$\pm$	0.16 	&				&				&		&	\\
130315A	&		&	0.30 	$\pm$	0.07 	&	1.87 	$\pm$	0.36 	&	4.61 	$\pm$	3.87 	&	3.71 	-	4.91 	&	0.81 	$\pm$	0.08 	&	1.07 	$\pm$	0.38 	&				&				&		&	\\
130609B	&		&	0.72 	$\pm$	0.08 	&	1.94 	$\pm$	0.05 	&	3.72 	$\pm$	2.92 	&	2.91 	-	5.45 	&	0.32 	$\pm$	0.04 	&	0.98 	$\pm$	0.23 	&				&				&		&	\\
140114A	&		&	0.22 	$\pm$	0.10 	&	1.27 	$\pm$	0.17 	&	4.33 	$\pm$	3.89 	&	3.08 	-	5.40 	&	1.06 	$\pm$	0.09 	&	1.16 	$\pm$	0.20 	&				&				&		&	\\
140323A	&		&	0.45 	$\pm$	0.08 	&	1.57 	$\pm$	0.07 	&	3.79 	$\pm$	3.19 	&	2.55 	-	4.94 	&	0.64 	$\pm$	0.04 	&	1.04 	$\pm$	0.18 	&				&				&		&	\\
140512A	&	0.725	&	0.74 	$\pm$	0.01 	&	1.65 	$\pm$	0.05 	&	4.16 	$\pm$	3.19 	&	2.51 	-	5.45 	&	0.45 	$\pm$	0.04 	&	0.76 	$\pm$	0.07 	&	2.14 	$\pm$	1.20 	&	2.22 	$\pm$	0.52 	&	0.70 	&	\\
140518A	&	4.707	&	0.15 	$\pm$	0.08 	&	1.64 	$\pm$	0.17 	&	3.46 	$\pm$	2.65 	&	2.54 	-	4.26 	&	0.97 	$\pm$	0.12 	&	1.01 	$\pm$	0.17 	&	3.57 	$\pm$	2.64 	&	2.67 	$\pm$	1.64 	&	1.18 	&	\\
140703A	&	3.14	&	0.60 	$\pm$	0.13 	&	2.35 	$\pm$	0.12 	&	4.15 	$\pm$	3.26 	&	3.61 	-	4.92 	&	0.84 	$\pm$	0.13 	&	0.71 	$\pm$	0.15 	&	3.35 	$\pm$	2.65 	&	2.94 	$\pm$	1.78 	&	1.40 	&	\\
140709A	&		&	0.55 	$\pm$	0.10 	&	1.28 	$\pm$	0.08 	&	4.13 	$\pm$	3.65 	&	3.30 	-	5.35 	&	0.74 	$\pm$	0.06 	&	0.81 	$\pm$	0.20 	&				&				&		&	\\
140818B	&		&	-0.18 	$\pm$	0.36 	&	1.17 	$\pm$	0.08 	&	3.41 	$\pm$	3.06 	&	2.49 	-	5.33 	&	0.99 	$\pm$	0.24 	&	1.29 	$\pm$	0.40 	&				&				&		&	\\
140916A	&		&	0.09 	$\pm$	0.03 	&	1.96 	$\pm$	0.11 	&	4.57 	$\pm$	3.46 	&	3.09 	-	5.53 	&	1.15 	$\pm$	0.26 	&	1.51 	$\pm$	0.15 	&				&				&		&	\\
141017A	&		&	0.11 	$\pm$	0.09 	&	1.14 	$\pm$	0.03 	&	3.37 	$\pm$	2.54 	&	2.49 	-	5.57 	&	0.66 	$\pm$	0.06 	&	1.01 	$\pm$	0.14 	&				&				&		&	\\
141031A	&		&	-0.04 	$\pm$	0.22 	&	1.05 	$\pm$	0.10 	&	4.35 	$\pm$	3.92 	&	3.93 	-	5.85 	&	0.31 	$\pm$	0.19 	&	0.73 	$\pm$	0.29 	&				&				&		&	\\
141121A	&	1.47	&	0.31 	$\pm$	0.09 	&	2.52 	$\pm$	0.23 	&	5.54 	$\pm$	4.59 	&	4.43 	-	6.10 	&	0.73 	$\pm$	0.13 	&	0.82 	$\pm$	0.19 	&	0.93 	$\pm$	0.13 	&	2.43 	$\pm$	1.26 	&	-0.34 	&	\\
150428B	&		&	0.13 	$\pm$	0.10 	&	0.95 	$\pm$	0.12 	&	4.25 	$\pm$	3.94 	&	3.02 	-	5.81 	&	0.23 	$\pm$	0.04 	&	0.66 	$\pm$	0.02 	&				&				&		&	\\
150626A	&		&	0.11 	$\pm$	0.12 	&	0.92 	$\pm$	0.14 	&	3.98 	$\pm$	3.66 	&	3.03 	-	4.83 	&	0.87 	$\pm$	0.10 	&	0.81 	$\pm$	0.28 	&				&				&		&	\\
150910A	&	1.359	&	0.40 	$\pm$	0.03 	&	2.13 	$\pm$	0.06 	&	3.68 	$\pm$	2.47 	&	2.33 	-	5.31 	&	0.42 	$\pm$	0.12 	&	0.54 	$\pm$	0.04 	&	3.36 	$\pm$	2.29 	&	2.27 	$\pm$	1.19 	&	1.31 	&	\\
151027A	&	0.81	&	0.01 	$\pm$	0.07 	&	1.67 	$\pm$	0.02 	&	3.57 	$\pm$	2.32 	&	2.82 	-	5.90 	&	1.11 	$\pm$	0.18 	&	1.35 	$\pm$	0.36 	&	3.10 	$\pm$	2.00 	&	2.10 	$\pm$	0.44 	&	0.89 	&	\\
160327A	&		&	0.00 	$\pm$	0.31 	&	1.41 	$\pm$	0.09 	&	3.35 	$\pm$	2.83 	&	2.69 	-	5.00 	&	0.72 	$\pm$	0.05 	&	1.04 	$\pm$	0.04 	&				&				&		&	\\
160607A	&		&	0.67 	$\pm$	0.02 	&	1.41 	$\pm$	0.03 	&	3.50 	$\pm$	2.58 	&	2.10 	-	6.06 	&	0.84 	$\pm$	0.10 	&	0.73 	$\pm$	0.31 	&				&				&		&	\\
160630A	&		&	0.22 	$\pm$	0.08 	&	1.18 	$\pm$	0.04 	&	3.25 	$\pm$	2.52 	&	2.09 	-	5.44 	&	0.37 	$\pm$	0.02 	&	0.82 	$\pm$	0.06 	&				&				&		&	\\
161117A	&	1.549	&	0.30 	$\pm$	0.07 	&	1.18 	$\pm$	0.03 	&	3.88 	$\pm$	3.10 	&	2.94 	-	6.08 	&	0.32 	$\pm$	0.14 	&	0.87 	$\pm$	0.26 	&	2.59 	$\pm$	1.75 	&	3.04 	$\pm$	0.00 	&	0.53 	&	\\
170113A	&	1.968	&	0.45 	$\pm$	0.06 	&	1.25 	$\pm$	0.03 	&	3.65 	$\pm$	2.92 	&	2.53 	-	5.94 	&	0.82 	$\pm$	0.00 	&	0.93 	$\pm$	0.11 	&	3.18 	$\pm$	2.45 	&	1.92 	$\pm$	0.75 	&	0.75 	&	\\
170202A	&	3.645	&	-0.06 	$\pm$	0.12 	&	1.17 	$\pm$	0.04 	&	3.39 	$\pm$	2.58 	&	2.58 	-	5.52 	&	0.57 	$\pm$	0.12 	&	0.81 	$\pm$	0.19 	&	3.85 	$\pm$	2.93 	&	2.96 	$\pm$	1.43 	&	1.09 	&	\\
170317A	&		&	0.65 	$\pm$	0.07 	&	1.54 	$\pm$	0.11 	&	3.59 	$\pm$	2.99 	&	2.44 	-	5.05 	&	0.68 	$\pm$	0.07 	&	1.21 	$\pm$	0.20 	&				&				&		&	\\
170607A	&		&	0.36 	$\pm$	0.04 	&	1.00 	$\pm$	0.03 	&	4.29 	$\pm$	3.64 	&	3.02 	-	6.14 	&	0.78 	$\pm$	0.10 	&	1.00 	$\pm$	0.11 	&				&				&		&	\\
171120A	&		&	0.41 	$\pm$	0.04 	&	2.22 	$\pm$	0.96 	&	5.02 	$\pm$	4.45 	&	3.60 	-	5.68 	&	0.67 	$\pm$	0.06 	&	0.88 	$\pm$	0.17 	&				&				&		&	\\
171205A	&	0.0368	&	-0.26 	$\pm$	0.21 	&	1.07 	$\pm$	0.07 	&	4.94 	$\pm$	4.27 	&	4.20 	-	6.45 	&	0.42 	$\pm$	0.14 	&	0.86 	$\pm$	0.16 	&	-2.47 	$\pm$	-3.33 	&	-0.93 	$\pm$	-2.04 	&	-2.82 	&	\\
171222A	&	2.409	&	0.03 	$\pm$	0.28 	&	0.76 	$\pm$	0.13 	&	4.53 	$\pm$	4.46 	&	3.74 	-	5.98 	&	1.07 	$\pm$	0.16 	&	1.11 	$\pm$	0.28 	&	1.69 	$\pm$	1.28 	&	2.54 	$\pm$	1.38 	&	0.19 	&	\\
180115A	&	2.487	&	0.63 	$\pm$	0.06 	&	1.36 	$\pm$	0.09 	&	3.91 	$\pm$	3.48 	&	2.55 	-	5.31 	&	0.67 	$\pm$	0.22 	&	0.95 	$\pm$	0.11 	&	2.49 	$\pm$	2.05 	&	2.00 	$\pm$	1.15 	&	0.91 	&	\\
180329B	&	1.998	&	0.34 	$\pm$	0.08 	&	1.48 	$\pm$	0.08 	&	3.78 	$\pm$	3.09 	&	2.70 	-	5.18 	&	0.93 	$\pm$	0.13 	&	0.90 	$\pm$	0.10 	&	2.58 	$\pm$	1.88 	&	2.52 	$\pm$	1.40 	&	0.77 	&	\\
180411A	&		&	0.49 	$\pm$	0.05 	&	1.56 	$\pm$	0.06 	&	4.06 	$\pm$	3.19 	&	2.72 	-	5.46 	&	0.45 	$\pm$	0.03 	&	0.94 	$\pm$	0.15 	&				&				&		&	\\

\hline				
\enddata
\tablenotetext{a}{$t_b$ is the break time of light curves from our fitting, and $\alpha_1$ and
$\alpha_2$ are the decay slopes before and after the break time.} \tablenotetext{b}{The start ($T_1$)
and end ($T_2$) time of our fitting by a smooth broken power-law.} \tablenotetext{c}{The spectral index
of magnetar jet and wind.}\tablenotetext{d}{The plateau luminosity of our fits.}\tablenotetext{e}{The
isotropic energy releases of the prompt gamma-ray and MD wind.}

\end{deluxetable}
\end{center}

\begin{deluxetable}{llllllllllll}
\tablewidth{0pt} \tabletypesize{\footnotesize} \tablecaption{The derived parameters of newly-born
magnetars for the GRBs in our sample }\tablenum{2} \tablehead{ \colhead{GRB}& \colhead{$P_{0}$ (ms)}&
\colhead{$B_{p}$ ($\times 10^{15}$ G) }& 
\colhead{$\varepsilon_{\rm
lim}$}& } \startdata
Typical-long GRBs \\
\hline
050315	&	1.55 	$\pm$	0.21 	&	0.47 	$\pm$	0.10 	&	2.6e-3	&	\\
050319	&	1.19 	$\pm$	0.23 	&	0.54 	$\pm$	0.16 	&	2.3e-3	&	\\
050802	&	2.05 	$\pm$	0.29 	&	1.90 	$\pm$	0.39 	&	1.4e-2	&	\\
050814	&	1.41 	$\pm$	0.31 	&	0.61 	$\pm$	0.20 	&	3.1e-3	&	\\
050822	&	3.72 	$\pm$	0.63 	&	2.18 	$\pm$	0.58 	&	2.9e-2	&	\\
050826	&	28.60 	$\pm$	5.40 	&	7.88 	$\pm$	2.22 	&	8.0e-1	&	\\
051016B	&	7.85 	$\pm$	1.41 	&	3.52 	$\pm$	0.97 	&	9.8e-2	&	\\
060202	&	1.54 	$\pm$	0.03 	&	3.34 	$\pm$	0.08 	&	1.8e-2	&	\\
060204B	&	2.43 	$\pm$	0.55 	&	2.24 	$\pm$	0.74 	&	1.9e-2	&	\\
060502A	&	3.03 	$\pm$	0.70 	&	1.44 	$\pm$	0.51 	&	1.6e-2	&	\\
060604	&	2.79 	$\pm$	0.70 	&	1.65 	$\pm$	0.63 	&	1.6e-2	&	\\
060605	&	1.58 	$\pm$	0.13 	&	1.65 	$\pm$	0.20 	&	9.3e-3	&	\\
060607A	&	0.58 	$\pm$	0.03 	&	0.47 	$\pm$	0.03 	&	9.6e-4	&	\\
060614	&	24.70 	$\pm$	1.55 	&	5.56 	$\pm$	0.51 	&	4.9e-1	&	\\
060714	&	2.14 	$\pm$	0.57 	&	3.41 	$\pm$	1.36 	&	2.6e-2	&	\\
060729	&	3.03 	$\pm$	0.12 	&	0.67 	$\pm$	0.04 	&	7.2e-3	&	\\
060814	&	1.80 	$\pm$	0.26 	&	1.29 	$\pm$	0.28 	&	8.2e-3	&	\\
061121	&	1.67 	$\pm$	0.18 	&	1.36 	$\pm$	0.21 	&	8.1e-3	&	\\
061202	&	0.98 	$\pm$	0.10 	&	0.61 	$\pm$	0.10 	&	2.1e-3	&	\\
061222A	&	0.83 	$\pm$	0.03 	&	0.84 	$\pm$	0.05 	&	2.5e-3	&	\\
070110	&	1.33 	$\pm$	0.04 	&	0.78 	$\pm$	0.03 	&	3.7e-3	&	\\
070129	&	2.48 	$\pm$	0.38 	&	1.44 	$\pm$	0.35 	&	1.3e-2	&	\\
070306	&	1.42 	$\pm$	0.10 	&	0.61 	$\pm$	0.06 	&	3.1e-3	&	\\
070328	&	0.95 	$\pm$	0.05 	&	2.86 	$\pm$	0.24 	&	9.7e-3	&	\\
070508	&	2.17 	$\pm$	0.13 	&	4.36 	$\pm$	0.40 	&	3.4e-2	&	\\
070521	&	1.17 	$\pm$	0.10 	&	2.06 	$\pm$	0.25 	&	8.5e-3	&	\\
080310	&	2.06 	$\pm$	0.19 	&	1.61 	$\pm$	0.22 	&	1.2e-2	&	\\
080430	&	5.89 	$\pm$	0.78 	&	2.15 	$\pm$	0.45 	&	4.5e-2	&	\\
080905B	&	1.15 	$\pm$	0.46 	&	1.49 	$\pm$	0.87 	&	6.1e-3	&	\\
081029	&	1.48 	$\pm$	0.14 	&	1.11 	$\pm$	0.15 	&	5.9e-3	&	\\
090404	&	1.34 	$\pm$	0.26 	&	0.97 	$\pm$	0.30 	&	4.6e-3	&	\\
090407	&	2.72 	$\pm$	0.31 	&	0.73 	$\pm$	0.13 	&	7.1e-3	&	\\
090516A	&	1.10 	$\pm$	0.19 	&	0.92 	$\pm$	0.23 	&	3.6e-3	&	\\
090529	&	3.91 	$\pm$	2.49 	&	1.78 	$\pm$	1.80 	&	2.5e-2	&	\\
090618	&	2.81 	$\pm$	0.19 	&	2.00 	$\pm$	0.20 	&	2.0e-2	&	\\
091018	&	3.80 	$\pm$	0.68 	&	9.90 	$\pm$	2.70 	&	1.3e-1	&	\\
091029	&	2.04 	$\pm$	0.27 	&	1.64 	$\pm$	0.33 	&	1.2e-2	&	\\
100302A	&	2.18 	$\pm$	1.22 	&	1.00 	$\pm$	0.88 	&	7.8e-3	&	\\
100418A	&	6.31 	$\pm$	0.84 	&	1.05 	$\pm$	0.21 	&	2.3e-2	&	\\
100425A	&	6.36 	$\pm$	2.99 	&	2.88 	$\pm$	2.13 	&	6.5e-2	&	\\
100615A	&	0.98 	$\pm$	0.18 	&	0.36 	$\pm$	0.10 	&	1.3e-3	&	\\
100704A	&	0.95 	$\pm$	0.20 	&	0.50 	$\pm$	0.16 	&	1.7e-3	&	\\
100814A	&	1.36 	$\pm$	0.08 	&	0.25 	$\pm$	0.02 	&	1.2e-3	&	\\
100901A	&	1.82 	$\pm$	0.12 	&	0.69 	$\pm$	0.07 	&	4.5e-3	&	\\
100906A	&	1.87 	$\pm$	0.26 	&	1.28 	$\pm$	0.25 	&	8.5e-3	&	\\
110213A	&	0.98 	$\pm$	0.05 	&	1.25 	$\pm$	0.10 	&	4.4e-3	&	\\
110808A	&	6.24 	$\pm$	3.99 	&	1.56 	$\pm$	1.57 	&	3.5e-2	&	\\
111008A	&	0.75 	$\pm$	0.15 	&	1.26 	$\pm$	0.38 	&	3.4e-3	&	\\
111228A	&	4.19 	$\pm$	0.54 	&	2.25 	$\pm$	0.45 	&	3.3e-2	&	\\
120118B	&	2.52 	$\pm$	1.17 	&	3.34 	$\pm$	2.44 	&	3.0e-2	&	\\
120422A	&	28.00 	$\pm$	9.14 	&	5.41 	$\pm$	1.40 	&	3.1e-1	&	\\
120521C	&	1.98 	$\pm$	0.29 	&	1.60 	$\pm$	0.35 	&	1.1e-2	&	\\
120811C	&	1.71 	$\pm$	0.89 	&	3.22 	$\pm$	2.58 	&	2.0e-2	&	\\
121027A	&	1.56 	$\pm$	0.40 	&	0.30 	$\pm$	0.11 	&	1.7e-3	&	\\
140512A	&	2.31 	$\pm$	0.26 	&	1.14 	$\pm$	0.19 	&	9.4e-3	&	\\
140518A	&	1.80 	$\pm$	0.24 	&	3.62 	$\pm$	0.77 	&	2.3e-2	&	\\
140703A	&	0.90 	$\pm$	0.15 	&	0.70 	$\pm$	0.16 	&	2.2e-3	&	\\
141121A	&	2.27 	$\pm$	0.31 	&	0.28 	$\pm$	0.05 	&	2.2e-3	&	\\
150910A	&	1.14 	$\pm$	0.08 	&	1.15 	$\pm$	0.12 	&	4.7e-3	&	\\
151027A	&	1.54 	$\pm$	0.10 	&	1.53 	$\pm$	0.15 	&	8.4e-3	&	\\
161117A	&	2.31 	$\pm$	0.36 	&	1.92 	$\pm$	0.46 	&	1.6e-2	&	\\
170113A	&	1.63 	$\pm$	0.30 	&	1.90 	$\pm$	0.52 	&	1.1e-2	&	\\
170202A	&	1.27 	$\pm$	0.17 	&	2.50 	$\pm$	0.54 	&	1.1e-2	&	\\
171205A	&	144.01 	$\pm$	25.37 	&	22.40 	$\pm$	6.34 	&	1.2e+1	&	\\
171222A	&	3.57 	$\pm$	2.21 	&	1.63 	$\pm$	1.70 	&	2.1e-2	&	\\
180115A	&	2.90 	$\pm$	1.06 	&	2.71 	$\pm$	1.49 	&	2.8e-2	&	\\
180329B	&	2.83 	$\pm$	0.28 	&	2.85 	$\pm$	0.28 	&	2.9e-2	&	\\
\hline												
UlGRBs \\												
\hline												
101225A	&	0.95 	$\pm$	0.09 	&	1.58 	$\pm$	0.24 	&	5.3e-3	&	\\
170714A	&	1.42 	$\pm$	0.08 	&	1.88 	$\pm$	0.20 	&	9.5e-3	&	\\
\enddata
\end{deluxetable}


\clearpage

\clearpage

\clearpage

\clearpage

\clearpage

\clearpage

\clearpage

\clearpage

\clearpage

\clearpage

\clearpage

\clearpage

\end{document}